\documentclass[letterpaper,english,reprint, aps]{revtex4-1}
\usepackage[T1]{fontenc}
\usepackage[latin9]{inputenc}
\usepackage{textcomp}
\usepackage{amsmath}
\usepackage{amssymb}
\usepackage{graphicx}
\usepackage{setspace}

\makeatletter

\newcommand*\LyXThinSpace{\,\hspace{0pt}}
\pdfpageheight\paperheight
\pdfpagewidth\paperwidth

\newcommand{\lyxdot}{.}

\makeatother

\usepackage{babel}
\begin{document}
\title{The reverse flow and amplification of heat in a quantum-dot system}
\author{Jianying Du}
\thanks{These authors contributed equally to this work.}
\author{Tong Fu}
\thanks{These authors contributed equally to this work.}
\author{Jingyi Chen}
\author{Shanhe Su}
\email{sushanhe@xmu.edu.cn}

\author{Jincan Chen}
\email{jcchen@xmu.edu.cn}

\address{Department of Physics, Xiamen University, Xiamen 361005, People\textquoteright s
Republic of China}
\begin{abstract}
\singlespacing{}We demonstrate that when a quantum dot is embedded between the two
reservoirs described by different statistical distribution functions,
the reverse flow and amplification of heat can be realized by regulating
the energy levels of the quantum dot and the chemical potentials of
two reservoirs. The reverse heat flow and amplification coefficient
of the quantum device are calculated. The novelty of this device is
that the reverse flow of heat does not need externally driving force
and this seemingly paradoxical phenomenon does not violate the laws
of thermodynamics. It is further expounded that the quantum device
has some practical applications. For example, the device can work
as a micro/nano cooler. Moreover, the performance characteristics
of the cooler are revealed for different distribution functions. The
coefficients of performance of the cooler operated at different conditions
are calculated and the optimum selection criteria of key parameters
are supplied.

\begin{description}
\item [{PACS~numbers}] 05.90. +m, 05.70. \textendash a, 03.65.\textendash w,
51.30. +i
\end{description}
\end{abstract}
\maketitle

\section{introduction}

Thermal management devices aim to flexibly regulate the heat flow
in a way similar to electronic devices controlling electrical current
\citep{key-1,key-2,key-3,key-4}. The fundamental modes of heat flow
controls include thermal diodes \citep{key-5,key-6,key-7}, regulators
\citep{key-8,key-9,key-10}, and switches \citep{key-11,key-12,key-13},
which have important applications in heating, cooling, and energy
conversion. Quantum dots (QDs) as perfect energy filters due to their
discrete electronic states are of significant interest for designing
thermal management devices \citep{key-14,key-15,key-16,key-17}. 

Inserting a QD into a semiconductor nanowire, Josefsson \textit{et
al}. demonstrated a quantum heat engine operating close to the thermodynamic
efficiency limit \citep{key-18,key-19}. Dutta \textit{et al}. made
a tunable heat valve gate by controlling the heat flow in a Kondo-correlated
single-quantum-dot transistor \citep{key-20,key-21}. Zhang \textit{et
al}. proposed a quantum thermal transistor based on three Coulomb-coupled
quantum dots \citep{key-22,key-23}. Jaliel \textit{et al}. experimentally
realized a resonant tunneling energy harvester by connecting two quantum
dots in series with a hot cavity \citep{key-24}. Considering a serial
double quantum dot coupled to two electron reservoirs, Dorsch \textit{et
al}. showed that phonon-assisted transports enable the conversion
of heat into electrical power in an energy harvester \citep{key-25}.
These existing researches have laid the foundation for the concept
design and experimental development of new quantum devices. Now, we
consider one simple novel quantum device, where a QD with a single
transition energy is embedded in the middle of two reservoirs described
by different statistical distribution functions. It will be proved
that such a device can transfer heat from a low temperature reservoir
to a high temperature reservoir without external driving force and
the amplification of heat flow can be realized in the transfer process.

The concrete contents of the paper are organized as follows: In Sec.
\uppercase\expandafter{\romannumeral2}, we establish the model of
a quantum-dot device, which may realize the reverse flow of heat.
By applying the master equation approach, the thermodynamic characteristics
of the quantum device operating between two Fermi or two Bose reservoirs
are revealed. In Sec. \uppercase\expandafter{\romannumeral3}, the
reverse flow and amplification of heat are realized by reasonably
adjusting the energy levels of the quantum dot and the chemical potentials
of two reservoirs. In Sec. \uppercase\expandafter{\romannumeral4},
it is expounded that the quantum device can work as a micro/nano cooler.
The coefficients of performance of the cooler operated at different
conditions are calculated and the optimum operation regions of the
cooler are determined. Finally, some meaningful conclusions are drawn.

\section{The model description of a single quantum-dot device}

\begin{figure}
\includegraphics[scale=0.6]{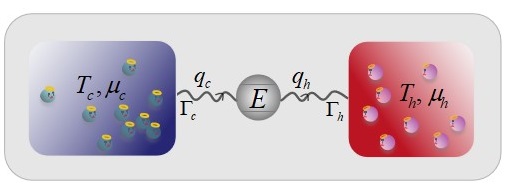}

\caption{The schematic diagram of a two-level QD embedded between the hot reservoir
$h$ and the cold reservoir $c$. \label{fig:1}}
\end{figure}

Figure \ref{fig:1} shows the model of a single quantum-dot system.
It is made up of a QD with two energy levels and two reservoirs described
by the distribution function

\begin{equation}
n_{\alpha}(E)=\frac{1}{\exp\left(\frac{E-\mu_{\alpha}}{k_{B}T_{\alpha}}\right)+\xi},\label{eq:1}
\end{equation}
where $T_{\alpha}(T_{h}>T_{c})$ and $\mu_{\alpha}$ are, respectively,
the temperature and chemical potential of reservoir $\alpha(=h,c)$,
$k_{B}$ is Boltzmann\textquoteright s constant, and $E$ is the transition
energy corresponding the energy difference of the empty and filled
states of the QD. For fermions, classical particles, and bosons, $\xi=1$,
$0$, and $-1$, respectively. $q_{c}$ is the heat flow flowing from
reservoir $c$ to the QD. $q_{h}$ is the heat flow flowing into reservoir
$h$ through the QD. $\Gamma_{a}$ is the coupling strength between
the QD and reservoir $\alpha$, which is a constant under the wideband
approximation \citep{key-26}. 

The free Hamiltonian of the QD in Fig.1 is given by

\begin{equation}
H=\frac{E}{2}\sigma^{z},\label{eq:2}
\end{equation}
where $\sigma^{z}$ denotes the component of the Pauli operator in
the z direction. The QD exchanges electrons with the nearby reservoirs
at energy level $E$. Let $\rho_{0}(\rho_{1})$ denote the probability
of finding the QD to be in empty (filled) state. Based on the Pauli
master equation \citep{key-26}, the dynamics equation governing the
evolution of the QD is given by

\begin{equation}
\frac{d}{dt}\left(\begin{array}{c}
\rho_{0}\\
\rho_{1}
\end{array}\right)=\left(\begin{array}{cc}
-P_{10}(E) & P_{01}(E)\\
P_{10}(E) & -P_{01}(E)
\end{array}\right)\left(\begin{array}{c}
\rho_{0}\\
\rho_{1}
\end{array}\right),\label{eq:3}
\end{equation}
where the effective transmission rates $P_{10}(E)=\sum_{\alpha=h,c}\Gamma_{a}n_{a}(E)$
and $P_{01}(E)=\sum_{\alpha=h,c}\Gamma_{a}\bar{n}_{\alpha}(E)$. For
a Fermi reservoir, $\bar{n}_{\alpha}(E)=1-n_{\alpha}(E)$. For a Bose
reservoir, $\bar{n}_{\alpha}(E)=1+n_{\alpha}(E)$. Note that $E$
is required to be larger than zero, because quantum coherence caused
by $E=0$ may influence the steady probability distribution so that
Eq. (\ref{eq:3}) loses its efficacy \citep{key-27}. Equation (\ref{eq:3})
has a very simple interpretation: the rate of a particle tunneling
into the QD from reservoir $\alpha$ is given by the bare tunneling
rate $\Gamma_{\alpha}$ multiplied by the distribution function $n_{\alpha}(E)$.
The rate of the inverse process, i.e., the tunneling of a particle
from the QD into reservoir $\alpha$, is given by the product of $\Gamma_{\alpha}$
and $\bar{n}_{\alpha}(E)$.

By using Eq. (\ref{eq:3}) and the relation $\rho_{0}+\rho_{1}=1$,
the particle current flowing from reservoir $c$ through the QD to
reservoir $h$ is \citep{key-17,key-28}

\begin{equation}
I_{M}^{c}=a_{0}\left[n_{c}(E)-n_{h}(E)\right],\label{eq:4}
\end{equation}
 where $a_{0}=\Gamma_{h}\Gamma_{c}/\left(\Gamma_{h}+\Gamma_{c}\right)$
for fermions and $a_{0}=\Gamma_{h}\Gamma_{c}/\left(\Gamma_{h}S_{h}+\Gamma_{c}S_{c}\right)$
with $S_{\alpha}=1+2n_{\alpha}(E)$ for bosons. Eq. (\ref{eq:4})
satisfies the relation $I_{M}^{c}=-I_{M}^{h}$, where $I_{M}^{h}$
is the particle current removing from reservoir $h$ through the QD
to reservoir $c$.

Each particle carries away an energy $E-\mu_{\alpha}$, when leaving
reservoir $\alpha$ through the QD \citep{key-29}. Thus, the heat
flows $q_{c}$ and $q_{h}$ are given by

\begin{equation}
q_{c}=\left(E-\mu_{c}\right)I_{M}^{c}\tag{5a}
\end{equation}
and 

\begin{equation}
q_{h}=\left(\mu_{h}-E\right)I_{M}^{h}\tag{5b},
\end{equation}
respectively. Equation (5) shows that the particle and heat flows
are proportional to each other. According to Eqs. (\ref{eq:4}) and
(5), $n_{c}(E)-n_{h}(E)$, which is the difference between the distribution
functions of reservoirs $c$ and $h$ at energy $E$, is one of the
main factors in determining the directions of the particle and heat
flows. At steady state, we are interested to reveal the mechanism
of the reverse heat flow of the device, which requires the heat flowing
out of the cold reservoir to be positive, i.e., $q_{c}>0$. 

The rate of entropy production $\dot{S}$ of the device can be obtained
by

\begin{equation}
\dot{S}=\frac{q_{h}}{T_{h}}-\frac{q_{c}}{T_{c}}=a_{0}\left(x_{c}-x_{h}\right)\left[n_{h}(E)-n_{c}(E)\right]\tag{6},\label{eq:6}
\end{equation}
where $x_{\alpha}=\left(E-\mu_{\alpha}\right)/T_{a}$. According the
second law of thermodynamics, $\dot{S}$ has to be not less than zero,
i.e., $\dot{S}>0$.

\begin{figure}
\includegraphics[scale=0.2]{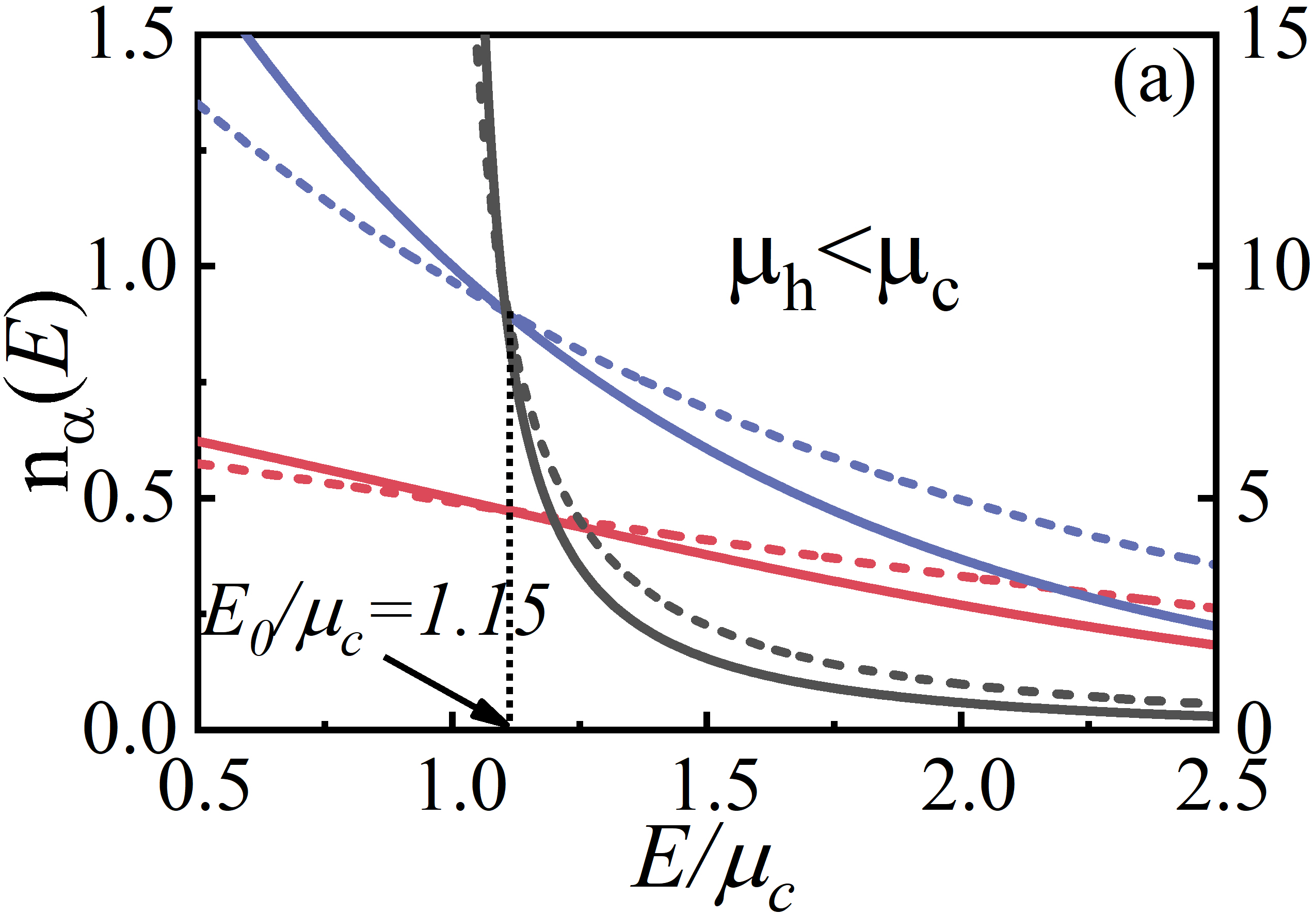}\includegraphics[scale=0.2]{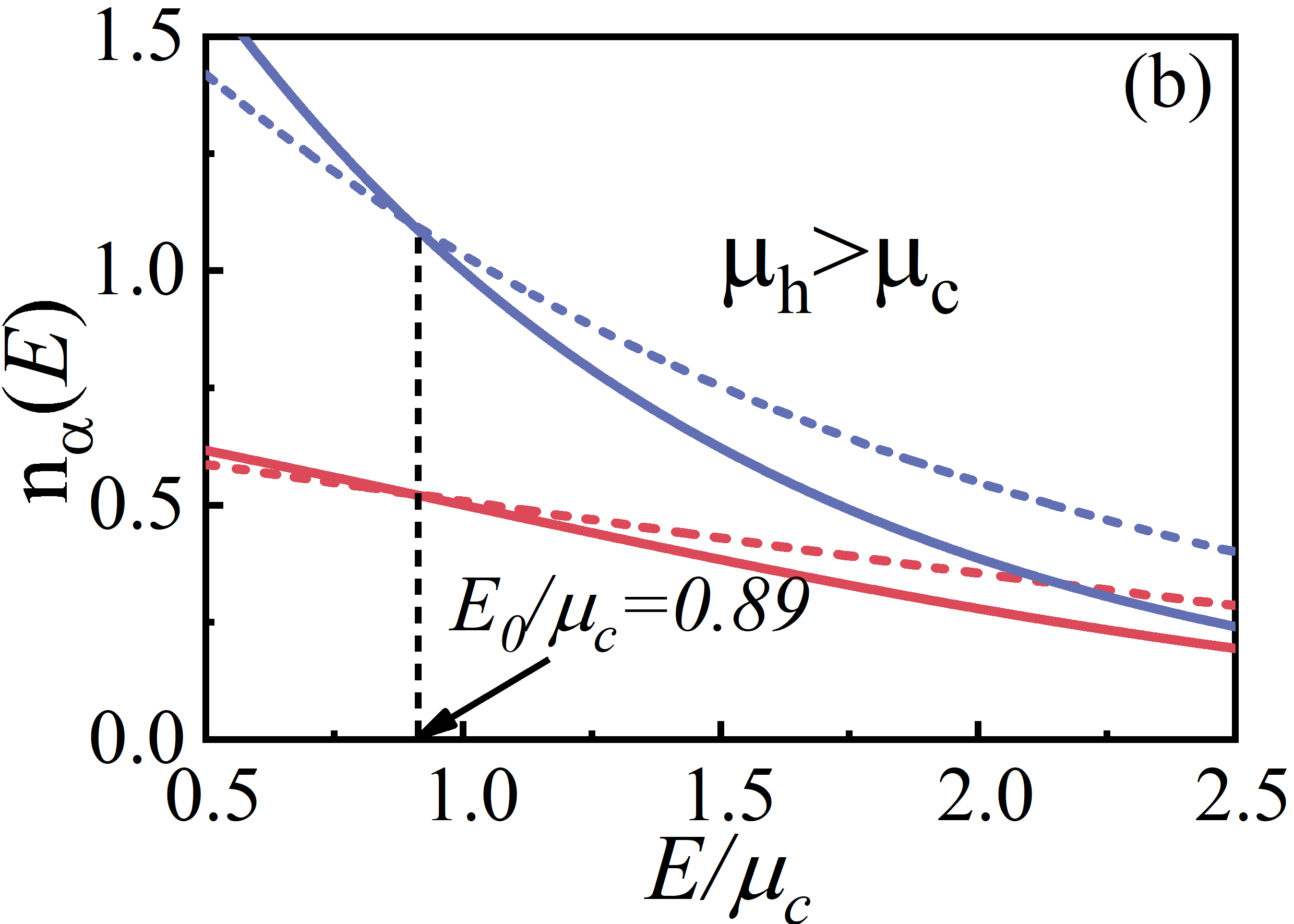}

\caption{The curves of different distribution functions varying with $E/\mu_{c}$.
The solid lines represent the mean occupation number $n_{c}(E)$ for
reservoir $c$ at low temperature, while the dashed lines are the
counterpart $n_{h}(E)$ for reservoir $h$ at high temperature. The
black, red, and blue lines indicate the Bose-Einstein, Fermi-Dirac,
and Maxwell\textendash Boltzmann functions, respectively. The left
vertical axis shows values for Fermi-Dirac and Maxwell\textendash Boltzmann
functions, while the corresponding scales of Bose-Einstein function
are on the right vertical axis. The mean occupation numbers of reservoirs
$h$ and $c$ at $E_{0}$ are equal, i.e., $n_{h}\left(E_{0}\right)=n_{c}\left(E_{0}\right)$.
The parameters $T_{h}=3$ and $T_{c}=2$, (a) $\mu_{h}=1.9$ and $\mu_{c}=2$,
and (b) $\mu_{h}=2$ and $\mu_{c}=1.9$ are chosen. In addition, Planck\textquoteright s
units are used by setting $\hbar=1$ and $k_{B}=1$ \citep{key-32,key-33}.\label{fig:2}}
\end{figure}

Using Eq. (\ref{eq:1}), one can plot the curves of the distribution
functions of the two reservoirs consisting of fermions, classical
particles, and bosons varying with the transport mode $E$, as shown
in Fig. \ref{fig:2}. For any given type of particles, the curves
of the distribution functions of reservoirs $c$ and $h$ intersect
at energy $E_{0}=\left(T_{h}\mu_{c}-\mu_{h}T_{c}\right)/\left(T_{h}-T_{c}\right)$
\citep{key-30,key-31}. Figures \ref{fig:2}$(a)$ and $(b)$ indicate,
respectively, the cases of $\mu_{h}<\mu_{c}$ and $\mu_{h}>\mu_{c}$.
For the case of $\mu_{h}>\mu_{c}$, $E_{0}<E<\mu_{c}$ and the distribution
functions of two Bose reservoirs are negative, which are not plotted
in Fig. \ref{fig:2}$(b)$. Therefore, the system consisting of a
single QD and two Bose reservoirs cannot realize the reverse flow
of heat for $\mu_{h}>\mu_{c}$. Such a case will not be discussed
below. Figure \ref{fig:2} shows that for different distribution functions,
$E_{0}/\mu_{c}$ has a same value, which depends directly on the temperatures
and chemical potentials. For example, for the parameter values given
in Fig. \ref{fig:2}, $E_{0}/\mu_{c}=1.15$ for the case of $\mu_{h}<\mu_{c}$
and $E_{0}/\mu_{c}=0.89$ for the case of $\mu_{h}>\mu_{c}$. Because
the Maxwell\textendash Boltzmann distribution function is a limit
case of the Fermi-Dirac distribution function, the results related
to the Maxwell\textendash Boltzmann distribution function may be directly
derived from those related to the Fermi-Dirac distribution function.
Thus, the case of the Maxwell\textendash Boltzmann distribution function
will be not discussed below.

It is seen from Eqs. (\ref{eq:4}) and (5) that when particles are
only transported at energy $E_{0}$, both the particle and heat flows
vanish, the rate of entropy production $\dot{S}=0$, and the device
can be operated reversibly. Note that $E=E_{0}$ is equivalent to
$x_{h}=x_{c}$. This shows that the condition of $E=E_{0}$ does not
require the temperatures and chemical potentials of two reservoirs
to be identical.

\begin{figure}
\includegraphics[scale=0.66]{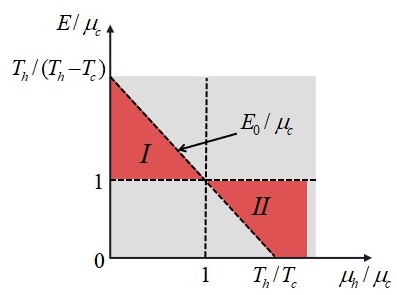}

\caption{The operative regions of the single QD device with a non-zero reverse
flow of heat. Region I is suitable for Fermi and Bose reservoirs.
Region II is only suitable for Fermi reservoirs. \label{fig:3}}
\end{figure}

According to the above analyses, we can further plot Fig. \ref{fig:3},
which determines the operative regions of the single QD with the reverse
flow of heat. In Fig. \ref{fig:3}, the vertical, horizontal, and
slash dashed lines represent the lines of $\mu_{h}/\mu_{c}=1$, $E_{0}/\mu_{c}=1$,
and $E_{0}/\mu_{c}$ within $\left[0,T_{h}/\left(T_{h}-T_{c}\right)\right]$,
respectively. Region I is bounded by the ordinate, the horizontal,
and slash dashed lines. In region I, $\mu_{h}<\mu_{c}$ , $\mu_{c}<E<E_{0}$,
and $n_{h}(E)<n_{c}(E)$, such that a positive particle current $I_{M}^{c}>0$
leaves the cold reservoir and there is a reverse flow of heat with
$q_{c}>0$. It shows that the single QD operates between two Fermi
(Bose) reservoirs with $\mu_{h}<\mu_{c}$, heat can be extracted from
the low-temperature reservoir. Region II is bounded by the three lines
of $E_{0}/\mu_{c}=1$, slash dashed line, and abscissa, while the
right margin of Region II will be determined below. In region II,
$\mu_{h}>\mu_{c}$ , $\mu_{c}>E>E_{0}$, and $n_{h}(E)>n_{c}(E)$,
such that a negative particle current $I_{M}^{c}<0$ leaves the low-temperature
reservoir and there is a reverse flow of heat $q_{c}>0$. It shows
that when the single QD operates between two Fermi reservoirs with
$\mu_{h}>\mu_{c}$, the heat of the low-temperature reservoir can
be also extracted.

Figure \ref{fig:3} implies the fact that when the parameters of the
QD and reservoirs satisfy the following relation

\[
\mu_{h}/\mu_{c}<1<E/\mu_{c}<E_{0}/\mu_{c}\tag{7a}
\]
or

\[
E_{0}/\mu_{c}<E/\mu_{c}<1<\mu_{h}/\mu_{c},\tag{7b}
\]
the device in Fig. \ref{fig:1} can have a non-zero reverse flow of
heat.

\section{The reverse flow and amplification of heat}

In the previous section, the necessary conditions for the reverse
flow of heat have been illustrated. It's important to choose the appropriate
transition energy of the QD and the temperatures and chemical potentials
of reservoirs. By using Eqs. (\ref{eq:4})-(\ref{eq:6}), the three-dimension
graphs of heat flow $q_{c}$, particle current $I_{M}^{c}$, and rate
of entropy production $\dot{S}$ of the single QD system operating
between two Fermi reservoirs varying with $E/\mu_{c}$ and $\mu_{h}/\mu_{c}$
are shown in Figs. 4(a)-(c), respectively. It is seen from Figs. 4(a)-(c)
that when $\mu_{h}/\mu_{c}<1<E/\mu_{c}<E_{0}/\mu_{c}$, $q_{c}>0$,
$I_{M}^{c}>0$, $\dot{S}>0$, and the device operates in region I
in Fig. \ref{fig:3}. This reveals the fact that an amount of heat
is released from the low temperature reservoir, accompanied by the
release of the particle current from the low temperature bath due
to the chemical potential difference.

\begin{figure}
\includegraphics[scale=0.1]{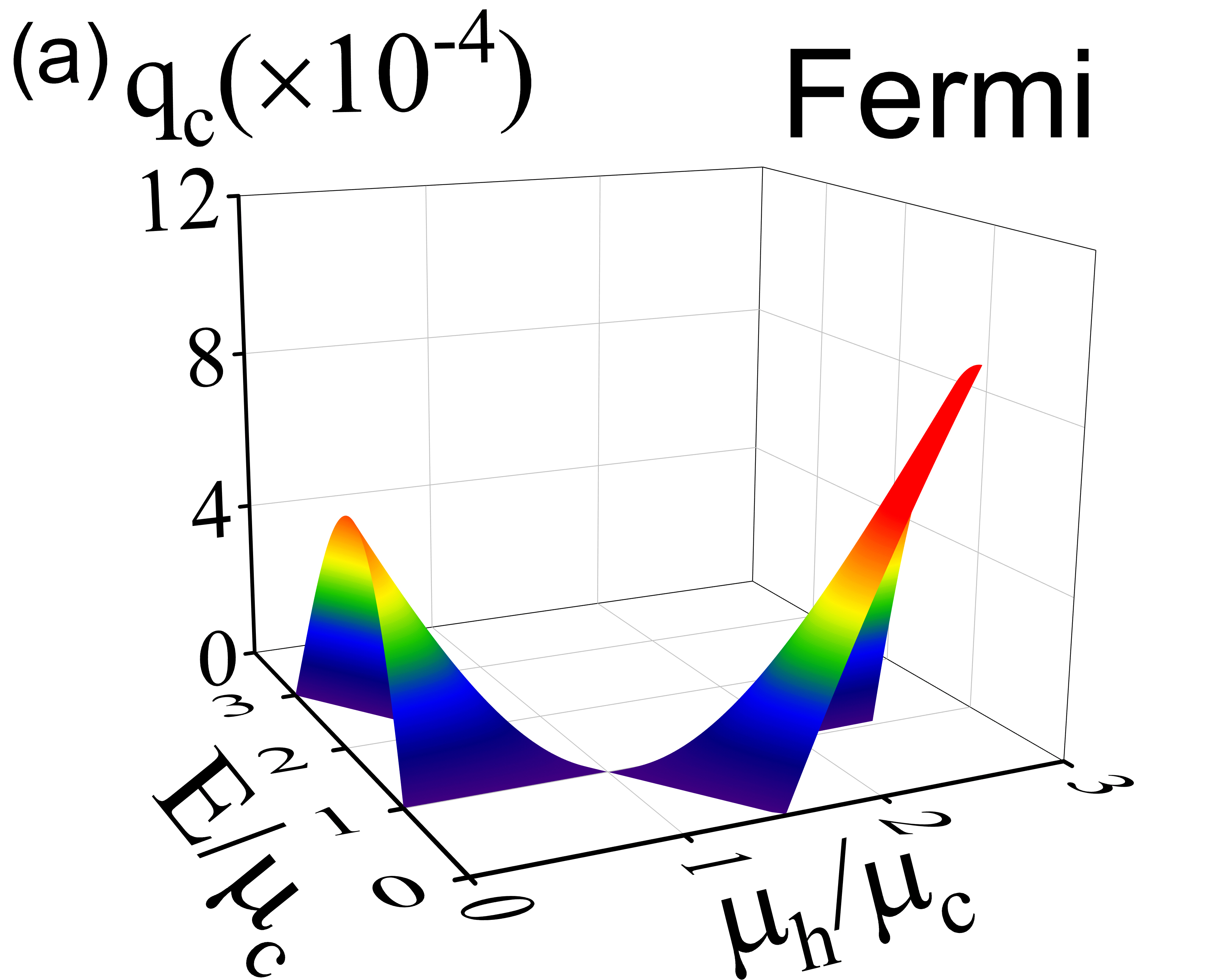}\includegraphics[scale=0.1]{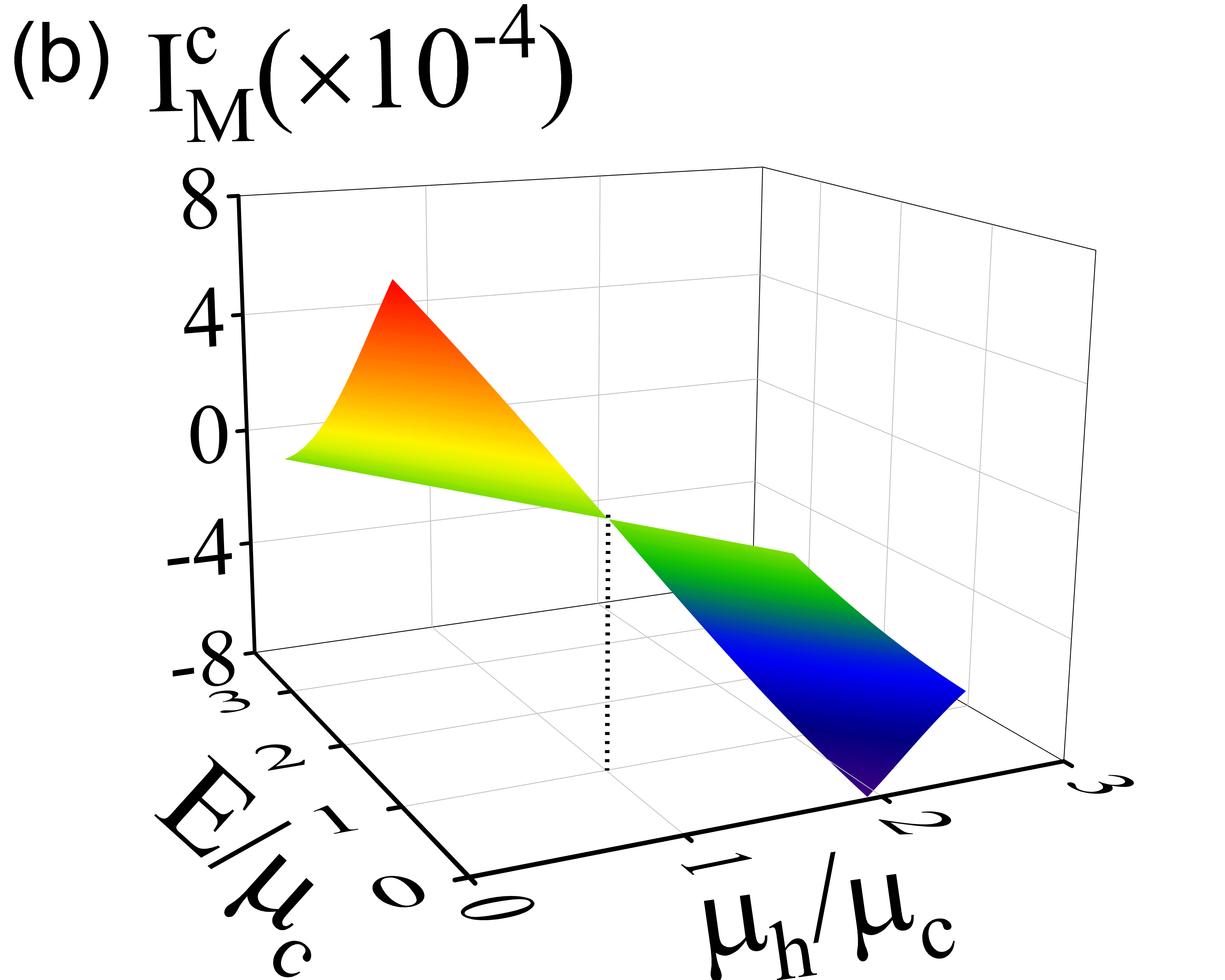}\includegraphics[scale=0.1]{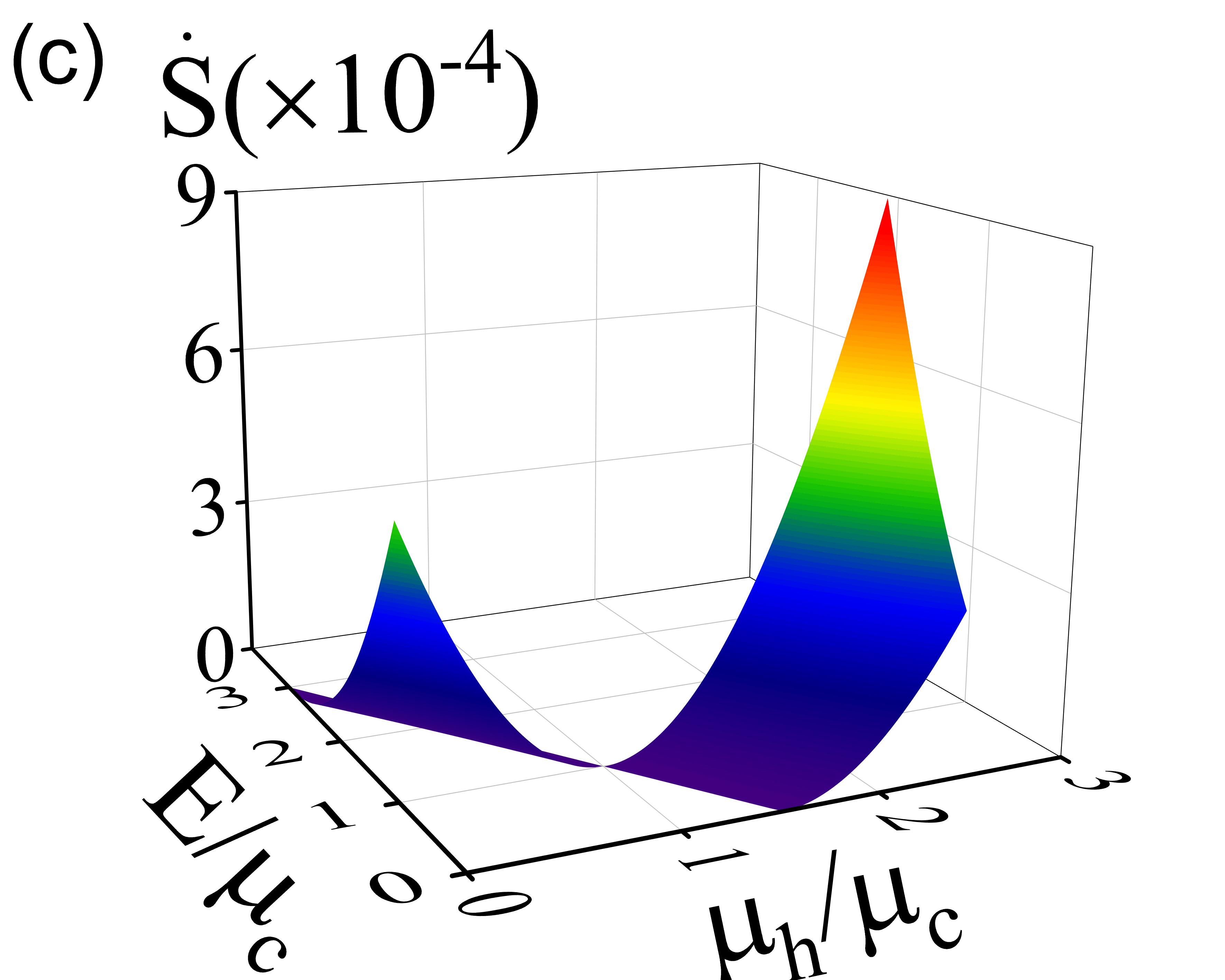}

\includegraphics[scale=0.1]{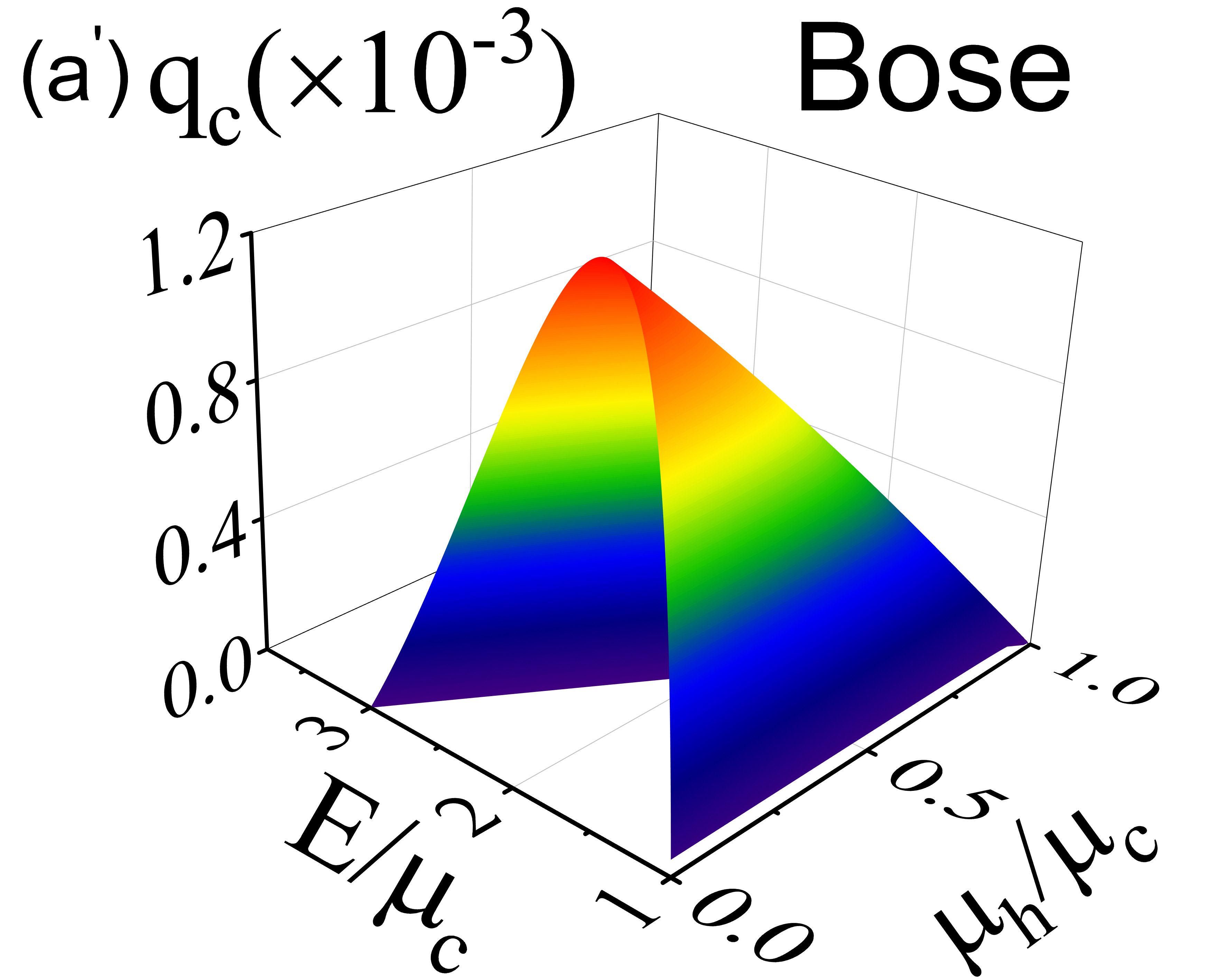}\includegraphics[scale=0.1]{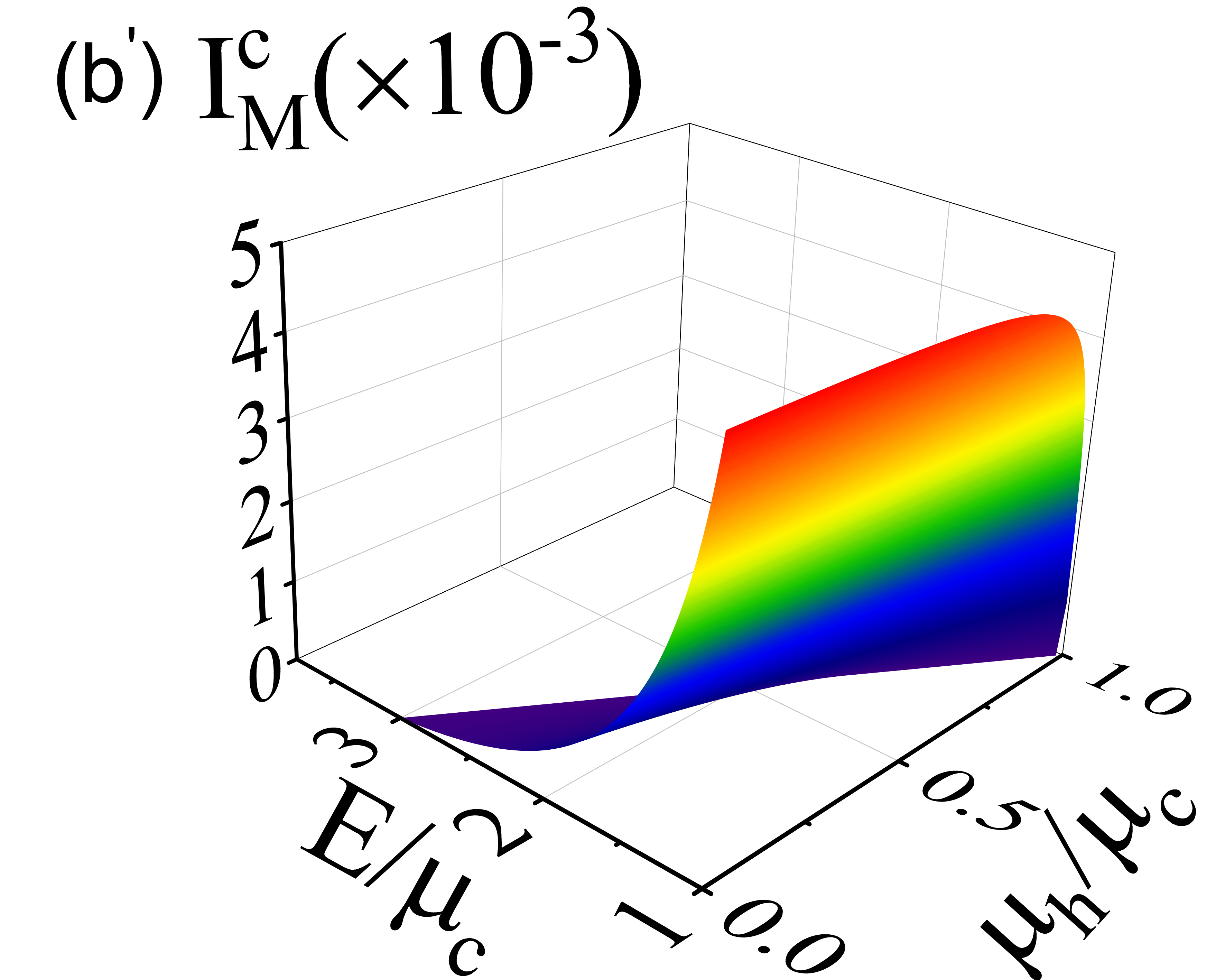}\includegraphics[scale=0.1]{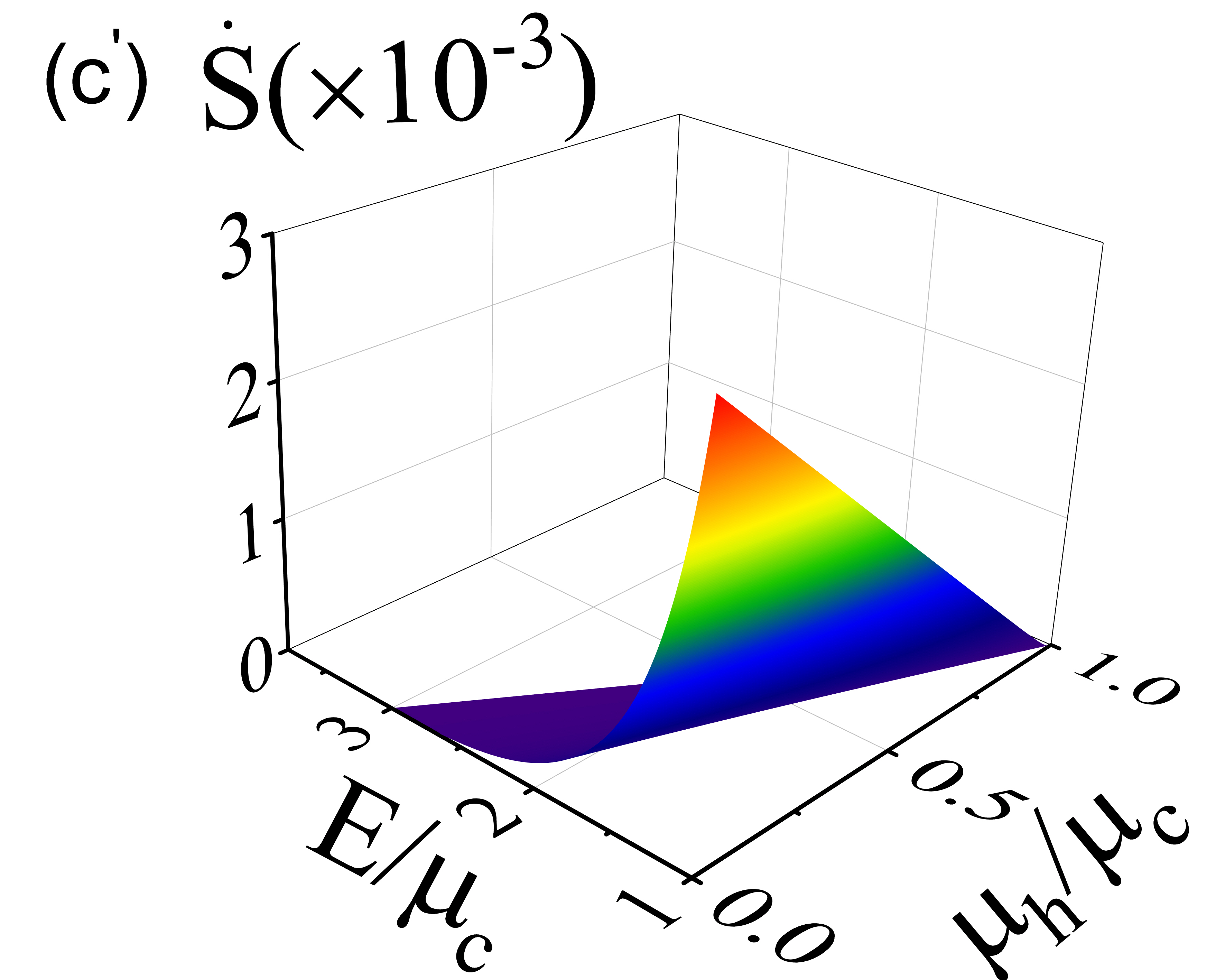}

\caption{Three-dimension graphs of ($a$) {[}($a'$){]} the heat flow $q_{c}$,
($b$) {[}($b'$){]} particle current $I_{M}^{c}$, and ($c$) {[}($c'$){]}
entropy production rate $\dot{S}$ of the QD device operating between
two Fermi (Bose) reservoirs as functions of $E/\mu_{c}$ and $\mu_{h}/\mu_{c}$.
The parameters $\mu_{c}=1.9$, $T_{h}=3$, $T_{c}=2$, and $\Gamma_{h}=\Gamma_{c}=0.01$.
\label{fig:4}}
\end{figure}

Similarly, when $E_{0}/\mu_{c}<E/\mu_{c}<1<\mu_{h}/\mu_{c}$, $q_{c}>0$,
$I_{M}^{c}<0$, $\dot{S}>0$, and the device operates in region II
in Fig. \ref{fig:3}. The releasing of heat from the cold reservoir
is accompanied by the particle current flowing into the cold reservoir.
It shows that the device operating in regions I and II in Fig. 3 have
a non-zero reverse flow of heat. It is found that $q_{c}$ is not
a monotonic function of $E/\mu_{c}$ in regions I and II. When $E=E_{0}$,
$q_{c}=0$ and $\dot{S}=0$, because the mean occupation numbers $n_{h}\left(E_{0}\right)=n_{c}\left(E_{0}\right)$
and the equal particles are transported reversibly between two reservoirs
$h$ and $c$. When $E=\mu_{c}\neq\mu_{h}$, $\quad Q_{c}=0$, but
$I_{M}^{c}\neq0$ and $\dot{S}>0$. Fig. \ref{fig:4}(a) shows that
there exists a local maximum of $q_{c}$ in regions I. However, in
region II, the maximum of $q_{c}$ increases with the increase of
$\mu_{h}/\mu_{c}$, but it also has an upper bound. This problem will
be discussed below. 

For two reservoirs described by the Bose-Einstein distribution function,
Eqs. (4)-(6) can be also used to generate the three dimensional graphs
of the heat flow $q_{c}$, particle current $I_{M}^{c}$, and rate
of entropy production $\dot{S}$ varying with $E/\mu_{c}$ and $\mu_{h}/\mu_{c}$,
as shown in Figs. 4($a'$)-($c'$), respectively. It is found from
Figs. 4($a'$)-($c'$) that only in the region of $\mu_{h}/\mu_{c}<1<E/\mu_{c}<E_{0}/\mu_{c}$,
can the device have a non-zero reverse flow of heat. Because the distribution
functions in reservoirs cannot be negative, there does not exist the
region of the reverse heat flow for $\mu_{h}>\mu_{c}$ when two reservoirs
are described by the Bose-Einstein distribution function. Thus, the
QD coupled with Bose reservoirs can work only in region I as a device
with a non-zero reverse flow of heat.

\begin{figure}
\includegraphics[scale=0.14]{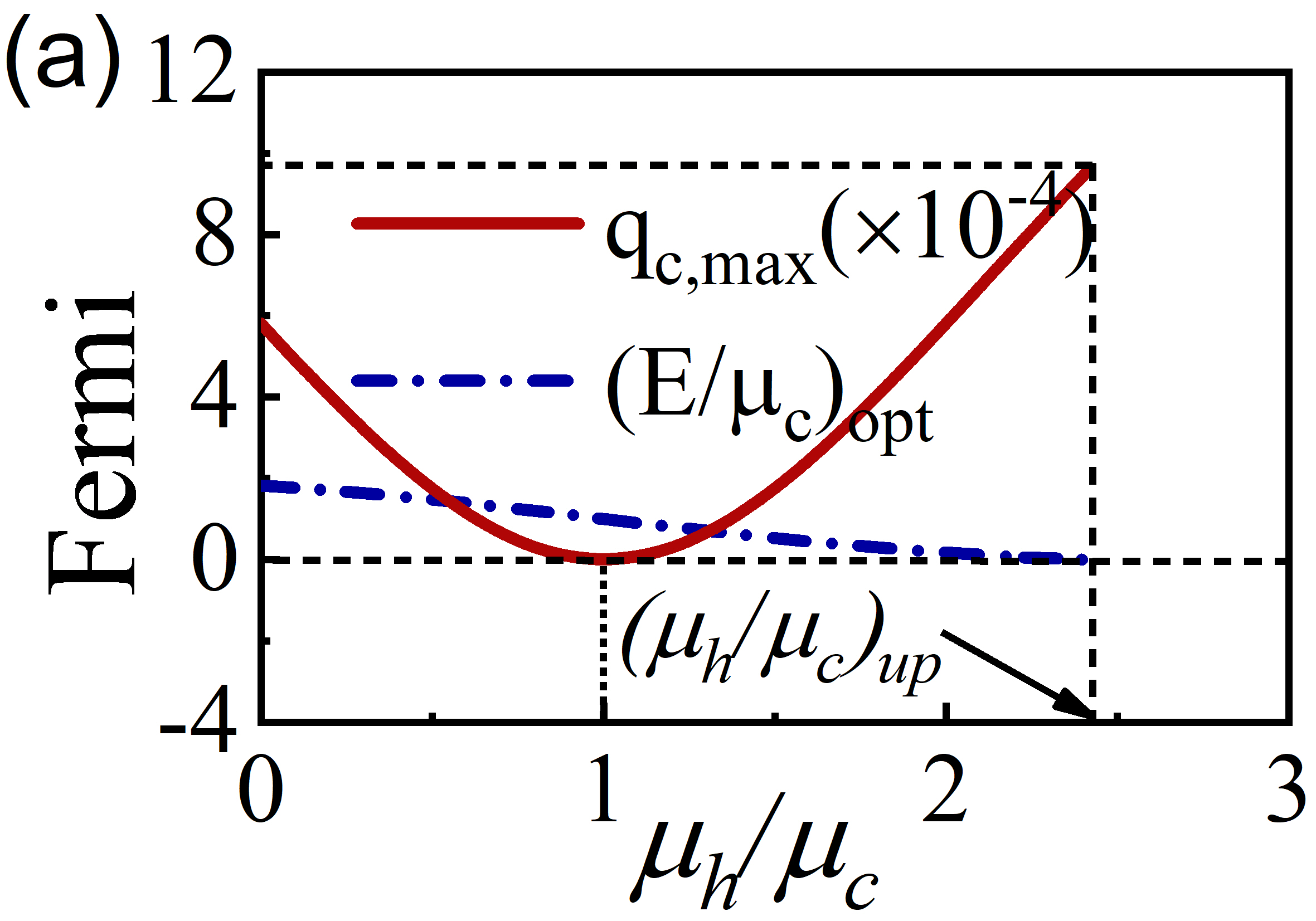}\includegraphics[scale=0.14]{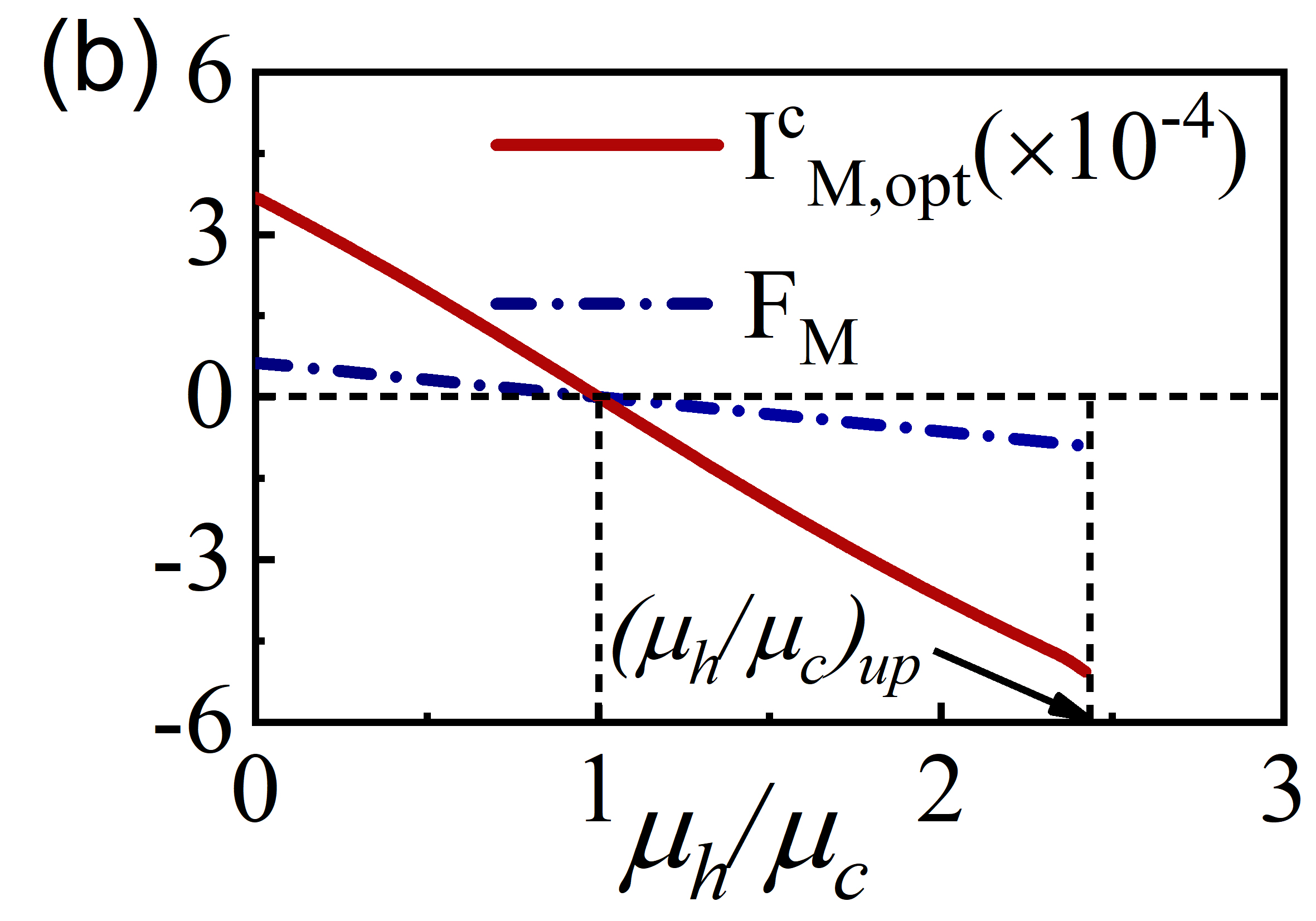}\includegraphics[scale=0.14]{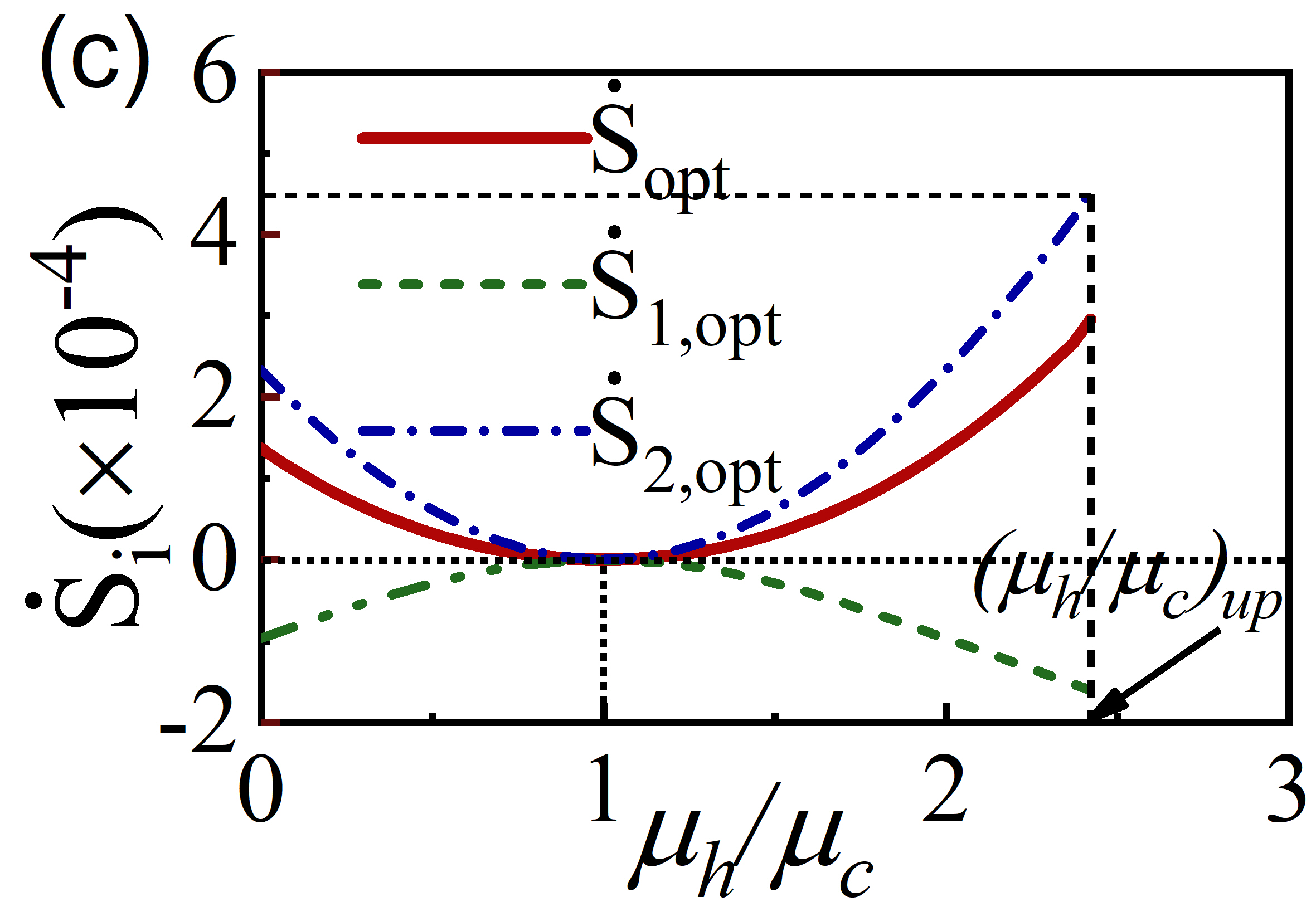}

\includegraphics[scale=0.14]{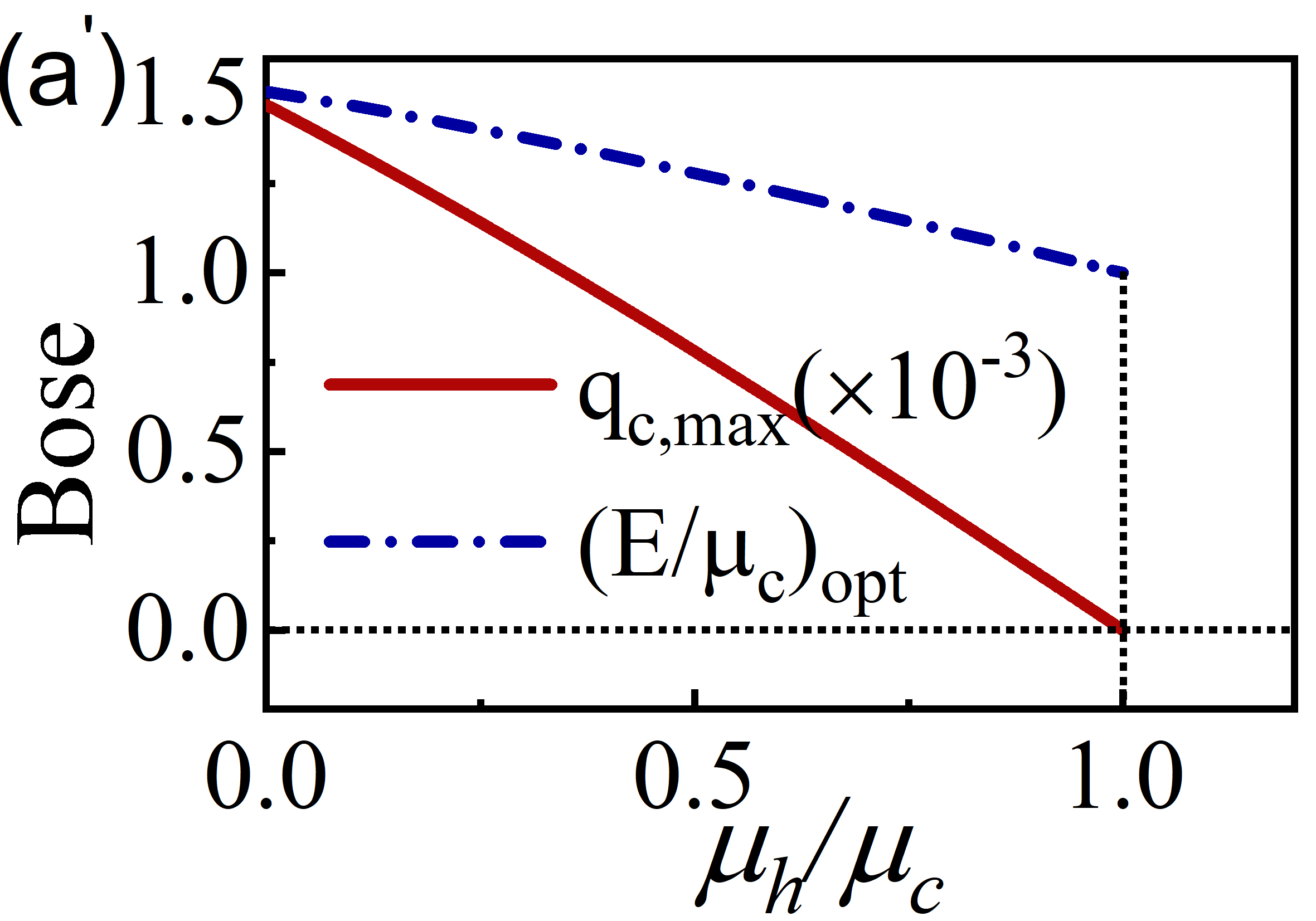}\includegraphics[scale=0.14]{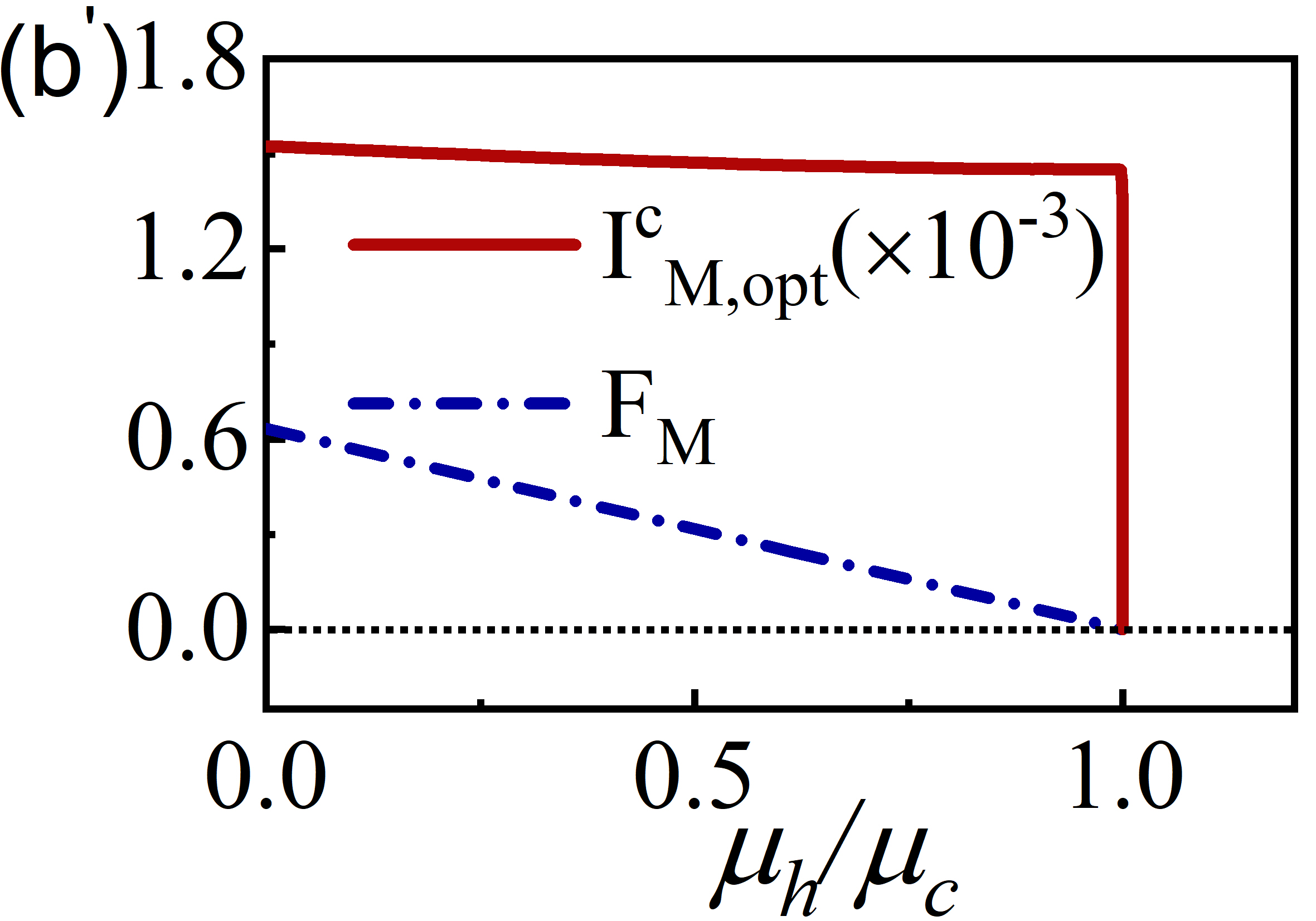}\includegraphics[scale=0.14]{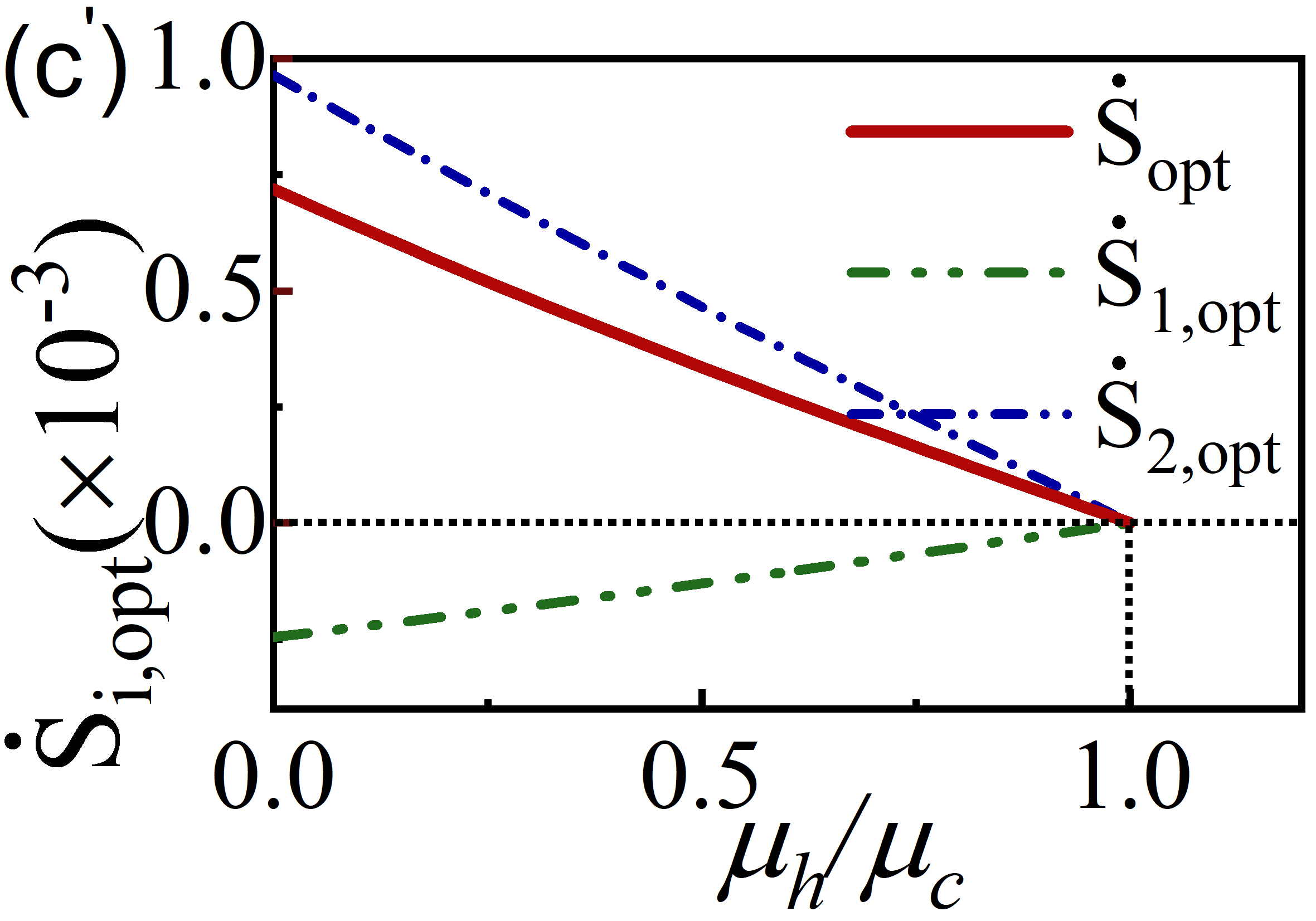}

\caption{($a$) {[}($a'$){]} The maximum heat flow $q_{c,max}$ and the optimized
energy level $(E/\mu_{c})_{opt}$, ($b$) {[}($b'$){]} the optimized
particle current $I_{M,opt}^{c}$ and $F_{M}$, and ($c$) {[}($c'$){]}
the optimized entropy production rate $\dot{S}_{i,opt},$ $(\dot{S}_{opt},\dot{S}_{1,opt},\dot{S}_{2,opt},)$
of the QD device operating between two Fermi {[}Bose{]} reservoirs
as functions of $\mu_{h}/\mu_{c}$. The parameters $\mu_{c}=1.9$,
$T_{h}=3$, $T_{c}=2$, and $\Gamma_{h}=\Gamma_{c}=0.01$. \label{fig:5}}
\end{figure}

It is seen clearly from Fig. \ref{fig:4} that for the parameter values
given by reservoirs, the heat flow $q_{c}$ is not a monotonic function
of the energy level $E$ of the QD. When the optimal value of the
energy level $E$ is chosen, i.e., $E/\mu_{c}=(E/\mu_{c})_{opt}$,
the heat flow $q_{c}$ attains its maximum, as indicated by Figs.
5($a$) {[}$(a')${]} for the QD device operating between two Fermi
(Bose) reservoirs. It is observed from Figs. \ref{fig:5}($a$) {[}($a'$){]}
that the maximum heat flow $q_{c,max}$ is a monotonically decreasing
function of $\mu_{h}/\mu_{c}$ in the region of $\mu_{h}/\mu_{c}<1$,
while it is a monotonically increasing function of $\mu_{h}/\mu_{c}$
in the region of $\mu_{h}/\mu_{c}>1$. $(E/\mu_{c})_{opt}$ is a monotonically
decreasing function of $\mu_{h}/\mu_{c}$. The condition of the model
availability requires $E>0$ so that $(E/\mu_{c})_{opt}$ has to be
larger than zero. When $(E/\mu_{c})_{opt}=0$, $\mu_{h}/\mu_{c}=(\mu_{h}/\mu_{c})_{up}$,
which is the upper bound of the ratio of two chemical potentials.
When $\mu_{h}/\mu_{c}\geq(\mu_{h}/\mu_{c})_{up}$, $(E/\mu_{c})_{opt}\leq0$
and the device can not have a reverse heat flow. Thus, $\mu_{h}/\mu_{c}=(\mu_{h}/\mu_{c})_{up}$
determines the right margin of region II in Fig. \ref{fig:3}. 

It is very interesting to note that the device in Fig. 1 has non-zero
reverse heat flows without externally driving force. This result does
not violate the second law of thermodynamics. The physical cause may
be explained as follows: In the system shown in Fig. 1, besides the
temperature difference, there exists another thermodynamic force resulting
from the chemical potential difference, which can be defined as $F_{M}=(\mu_{c}-\mu_{h})/T_{h}$.
This force is just the driving force of the reverse heat flow. In
order to further expound this problem, the rate of entropy production
may be rewritten as

\begin{equation}
\dot{S}=F_{T}q_{c}+F_{M}I_{M}^{c}=\dot{S}_{1}+\dot{S}_{2}\tag{8},\label{eq:8}
\end{equation}
where $\dot{S}_{1}=F_{T}q_{c}$ is the rate of entropy production
caused by the heat flow $q_{c}$, $F_{T}=1/T_{h}-1/T_{c}<0$ so that
$\dot{S}_{1}\leq0$, and $\dot{S}_{2}=F_{M}I_{M}^{c}$ is the rate
of entropy production caused by particle current. Using Eqs. (4),
(5), and (8), one can plot curves of the optimal values $I_{M,opt}^{c}$,
$\dot{S}_{opt}$, $\dot{S}_{1,opt}$, and $\dot{S}_{2,opt}$ of $I_{M}^{c}$,
$\dot{S}$, $\dot{S}_{1}$, and $\dot{S}_{2}$ at the maximum heat
flow varying with $\mu_{h}/\mu_{c}$, as indicated by Figs. 5($b$)
{[}($b'$){]} and ($c$) {[}($c'$){]} for the QD device operating
between two Fermi (Bose) reservoirs, where the $\mu_{h}/\mu_{c}\sim F_{M}$
curves are also given in Figs. \ref{fig:5}($b$) {[}($b'$){]}. It
is seen from Figs. \ref{fig:5}($b$) {[}($b'$){]} that the positive
or negative values of $I_{M,opt}^{c}$ directly depend on that of
$F_{M}$, so that $\dot{S}_{2}\geq0$. $I_{M,opt}^{c}$ is proportional
to $F_{M},$ and both $I_{M,opt}^{c}$ and $F_{M}$ are of monotonically
decreasing functions of $\mu_{h}/\mu_{c}$. When $\mu_{h}/\mu_{c}=1$,
both $I_{M,opt}^{c}$ and $F_{M}$ are equal to zero. It is observed
from Figs. \ref{fig:5}($c$) {[}($c'$){]} that $\dot{S}\geq0$ because
$\dot{S}_{2}\geq|\dot{S}_{1}|$. It shows once again that this seemingly
paradoxical phenomenon appearing in the QD device does not violate
the second law of thermodynamics. 

It is seen from Fig. \ref{fig:5} that in the region of $\mu_{h}/\mu_{c}>1$,
the heat flow $q_{c}$ of the QD device operating between two Fermi
reservoirs attains its maximum when $\mu_{h}/\mu_{c}\rightarrow(\mu_{h}/\mu_{c})_{up}$;
while in the region of $\mu_{h}/\mu_{c}<1$, the heat flow of the
QD device obtains its maximum when $\mu_{h}/\mu_{c}=0$, and the maximum
heat flow of the QD device operating between two Bose reservoirs is
larger than that operating between two Fermi reservoirs

\begin{figure}
\includegraphics[scale=0.12]{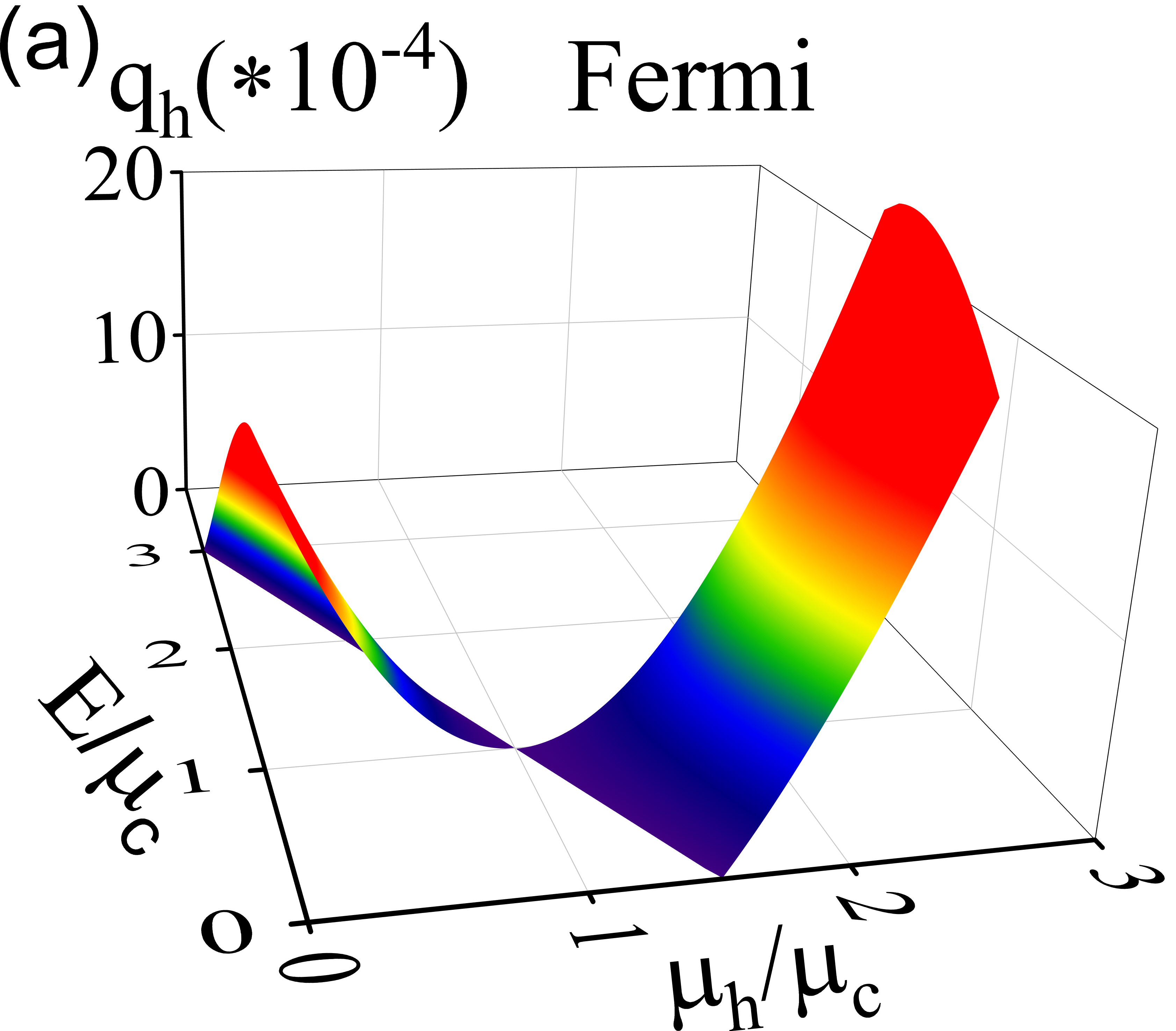}\includegraphics[scale=0.12]{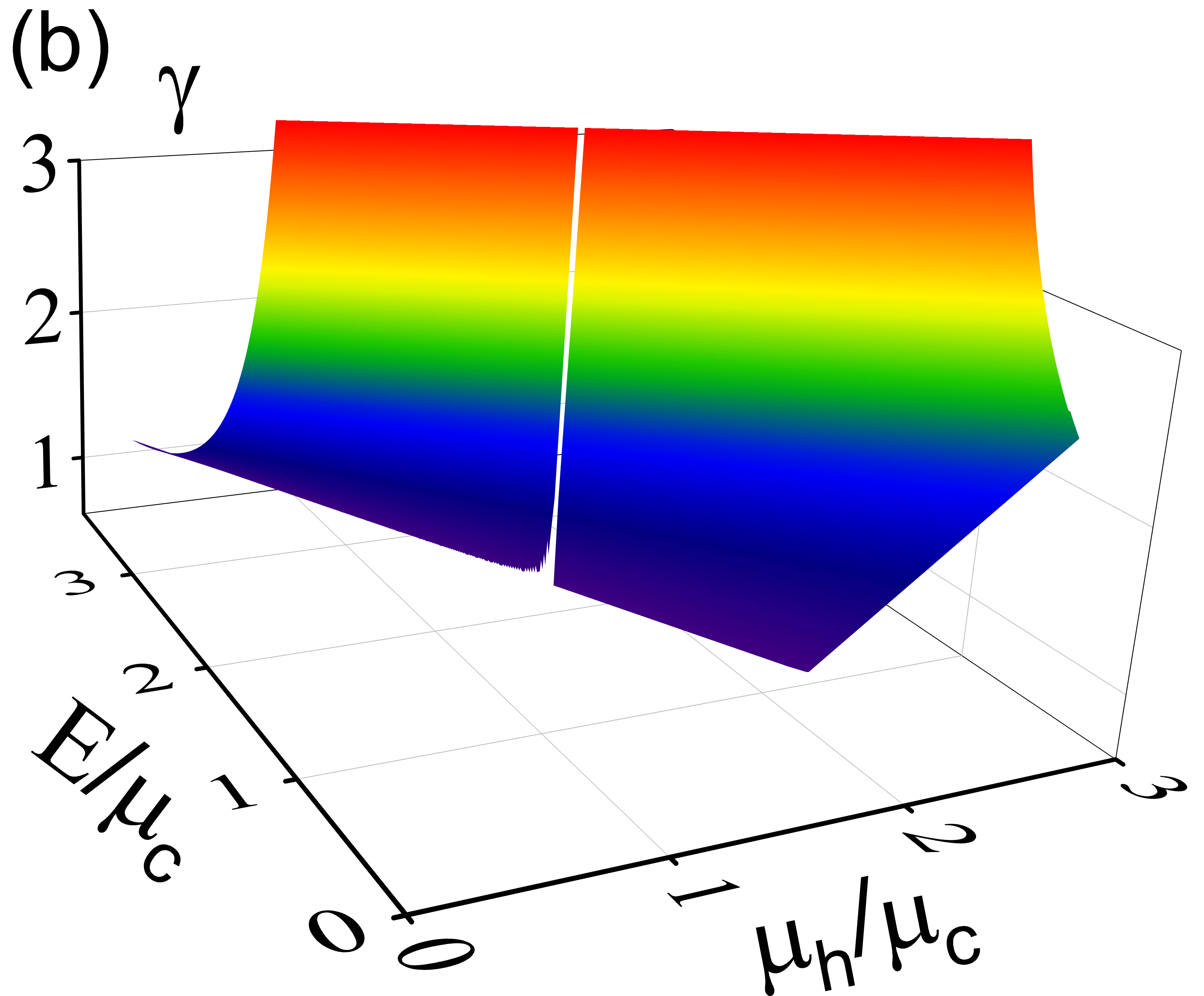}

\includegraphics[scale=0.12]{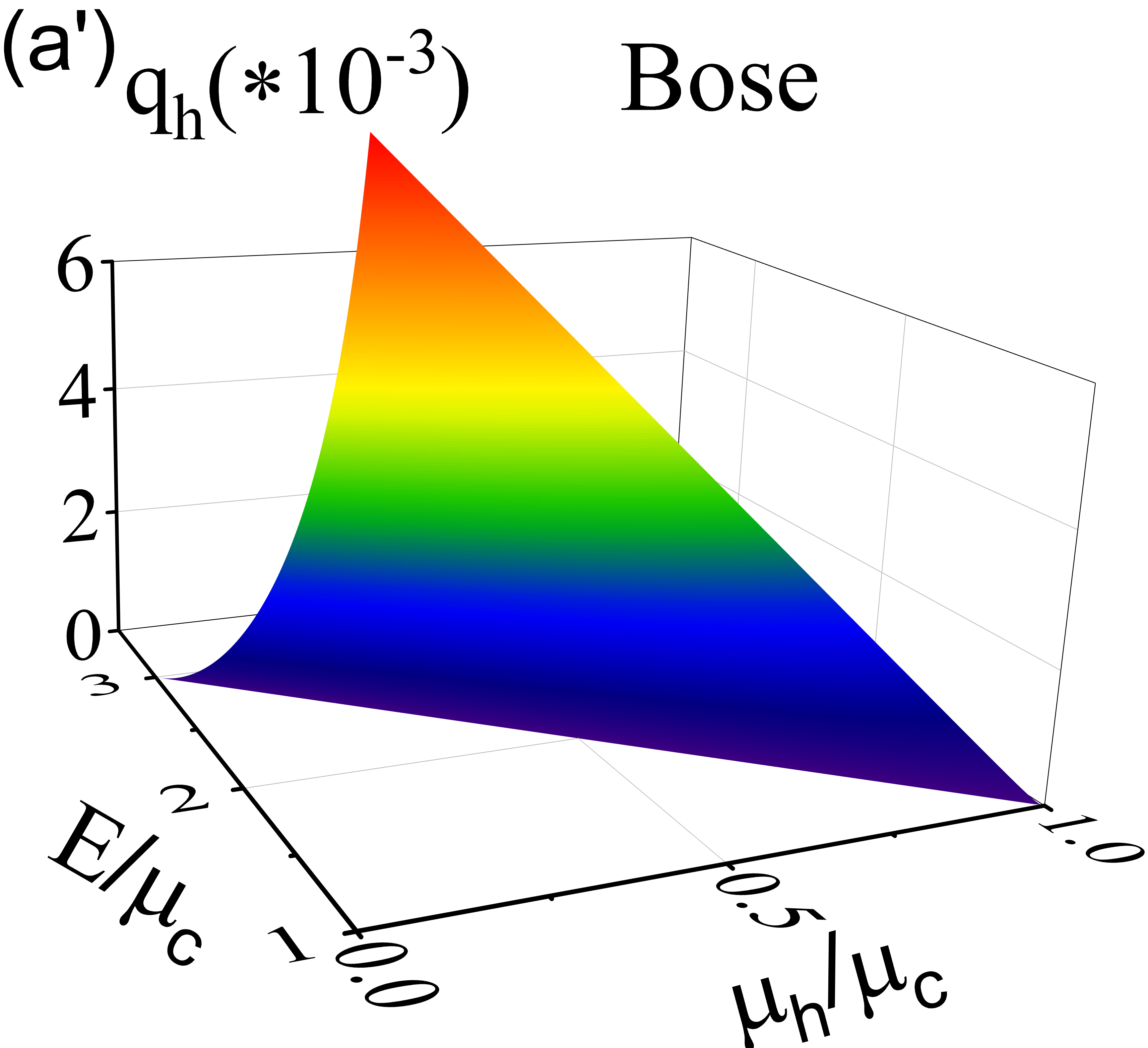}\includegraphics[scale=0.12]{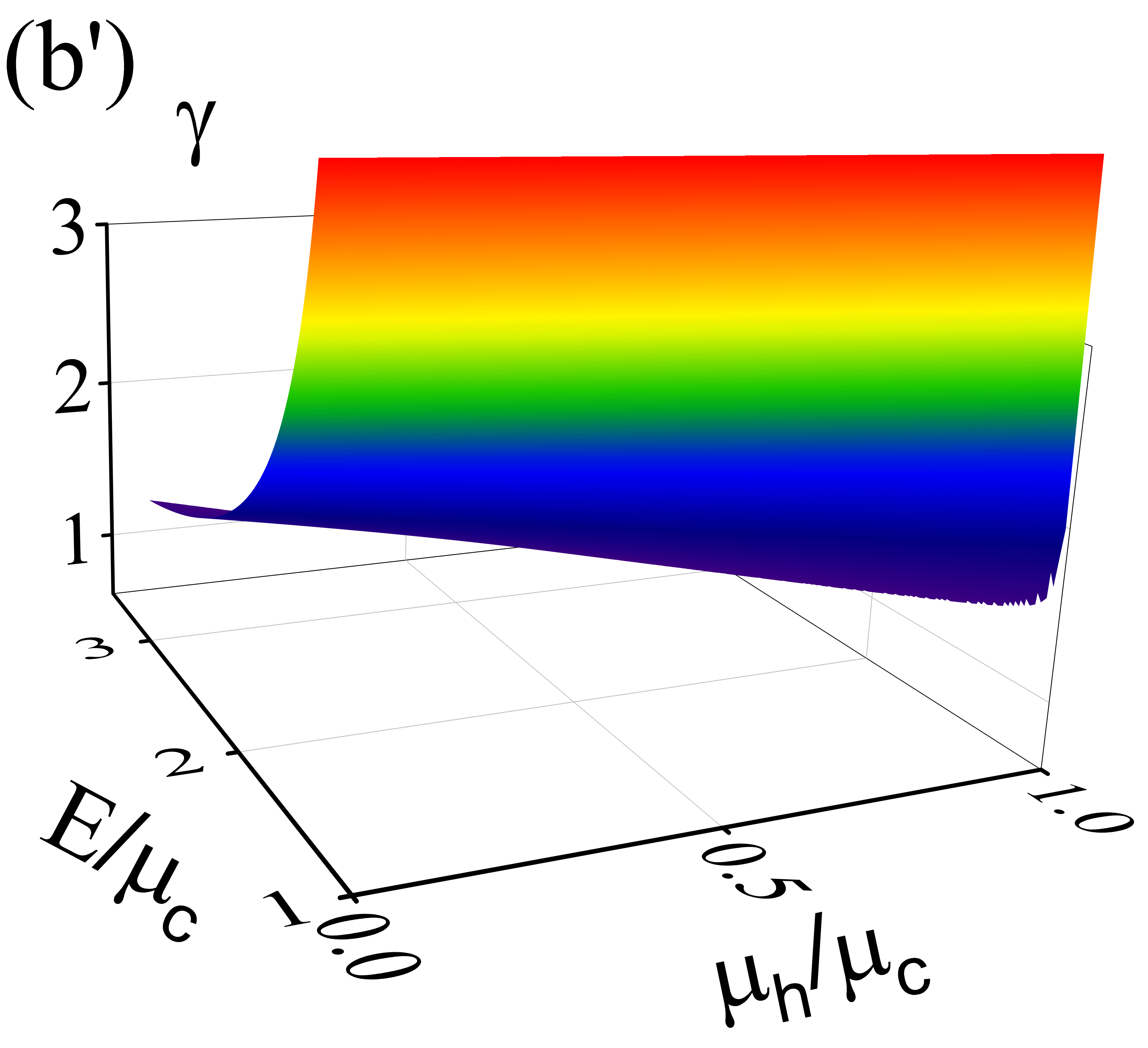}

\caption{Three-dimension graphs of ($a$) {[}($a'$){]} the other heat flow
$q_{h}$, ($b$) {[}($b'$){]} the amplification coefficient $\gamma$
of the QD device operating between two Fermi (Bose) reservoirs as
functions of $E/\mu_{c}$ and $\mu_{h}/\mu_{c}$. The parameters $\mu_{c}=1.9$,
$T_{h}=3$, $T_{c}=2$, and $\Gamma_{h}=\Gamma_{c}=0.01$. \label{fig:6}}
\end{figure}

Using Eqs. (4) and (5), we can also plot the three-dimension graphs
of another heat flow $q_{h}$ of the single QD device operating between
two Fermi (Bose) reservoirs varying with $E/\mu_{c}$ and $\mu_{h}/\mu_{c}$,
as shown in Fig. \ref{fig:6}($a$) {[}($a'$){]}. It is observed
from Fig. \ref{fig:6} that in regions I and II shown in Fig. \ref{fig:3},
$q_{h}>q_{c}$. This implies the fact that the heat flow is amplified
in the reverse transfer process of heat. It is seen from Eq. (5) that
in the reverse transfer process of heat, the increment of heat is
caused by the reduction of particle energy flow. It indicates that
the amplification of heat does not violate the first law of thermodynamics.
If defining an amplification coefficient as $\gamma=q_{h}/q_{c}$,
one can plot the three-dimension graphs of the amplification coefficient
$\gamma$ of the single QD device operating between two Fermi ( Bose)
reservoirs varying with $E/\mu_{c}$ and $\mu_{h}/\mu_{c}$, as shown
in Fig. \ref{fig:6}($b$) {[}($b'$){]}. It can be observed that
when $E=E_{0}$, $q_{c}=0$, $q_{h}=0$, and $\gamma=T_{h}/T_{c}>1$.
When $E=\mu_{c}\neq\mu_{h}$, $\gamma\rightarrow\infty$. In regions
I and II shown in Fig. \ref{fig:3}, $q_{h}>q_{c}$ and $\gamma>T_{h}/T_{c}$.

\section{An instance of the QD device}

It is note worthy that the QD device can work as a micro/nano cooler
because heat can be extracted from the cold reservoir. Using Eq. (5),
one can calculate the coefficient of performance of the QD cooler
as

\begin{equation}
\varepsilon=q_{c}/(q_{h}-q_{c})=\frac{E/\mu_{c}-1}{1-\mu_{h}/\mu_{c}}.\tag{9}\label{eq:9}
\end{equation}
\begin{figure}
\includegraphics[scale=0.15]{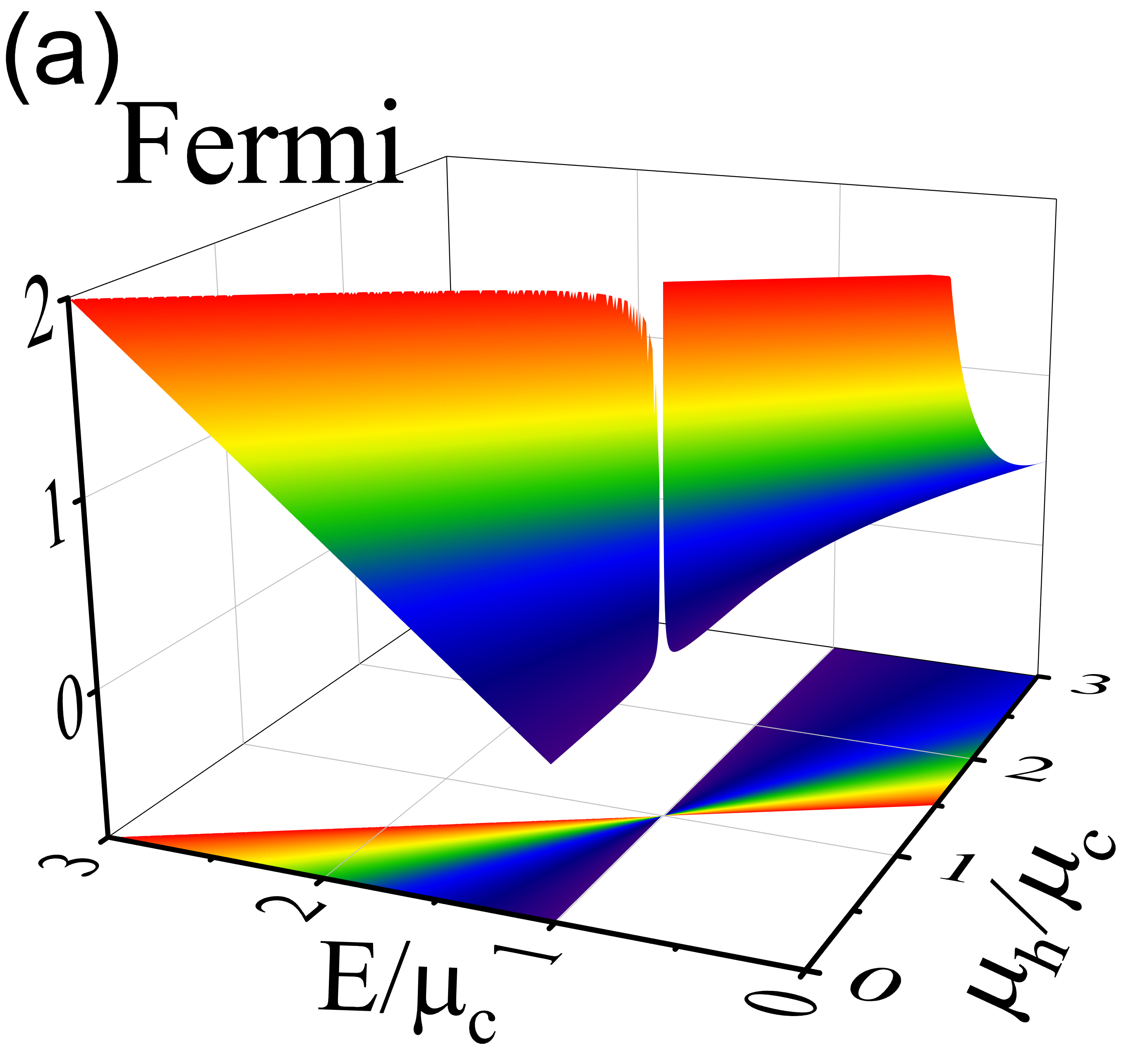}\includegraphics[scale=0.15]{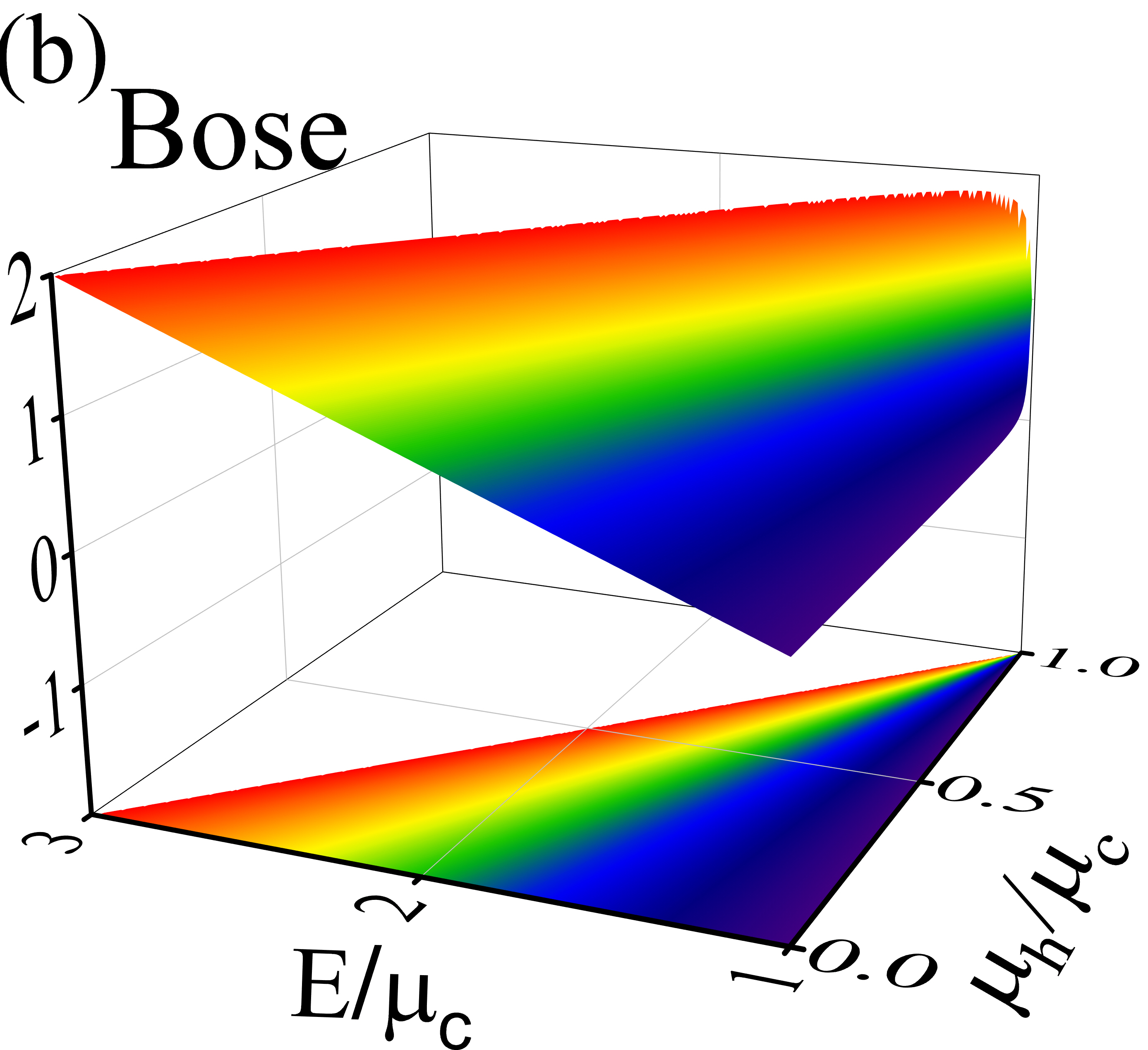}

\caption{Three-dimension graphs of ($a$) {[}($b$){]} the coefficient of performance
$\varepsilon$ of the QD cooler operating between two Fermi (Bose)
reservoirs as functions of $E/\mu_{c}$ and $\mu_{h}/\mu_{c}$. The
parameters $\mu_{c}=1.9$, $T_{h}=3$, $T_{c}=2$, and $\Gamma_{h}=\Gamma_{c}=0.01$.
\label{fig:7}}
\end{figure}
Equation (9) indicates that in region I shown in Fig. \ref{fig:3},
$\varepsilon$ increases monotonically with the increase of $E/\mu_{c}$
and $\mu_{h}/\mu_{c}$; while in region II shown in Fig. \ref{fig:3},
$\varepsilon$ decreases monotonically with the increase of $E/\mu_{c}$
and $\mu_{h}/\mu_{c}$; as illustrated in Fig. \ref{fig:7}. In order
to see the change trend of $\varepsilon$ more clearly, the projection
graphs of $\varepsilon$ are also plotted in Fig. \ref{fig:7}. When
$E=E_{0}$, $q_{c}=0$ and $\varepsilon=T_{c}/(T_{h}-T_{c})\equiv\varepsilon_{r}$,
which is the reversible coefficient of performance of the Carnot refrigerator
operating between the two heat reservoirs at temperatures $T_{h}$
and $T_{c}$. When $E=\mu_{c}\neq\mu_{h}$, $q_{c}=0$ and $\varepsilon=0$.
In regions I and II shown in Fig. \ref{fig:3}, $q_{c}>0$ and $0<\varepsilon<\varepsilon_{r}$.
\begin{figure}
\includegraphics[scale=0.12]{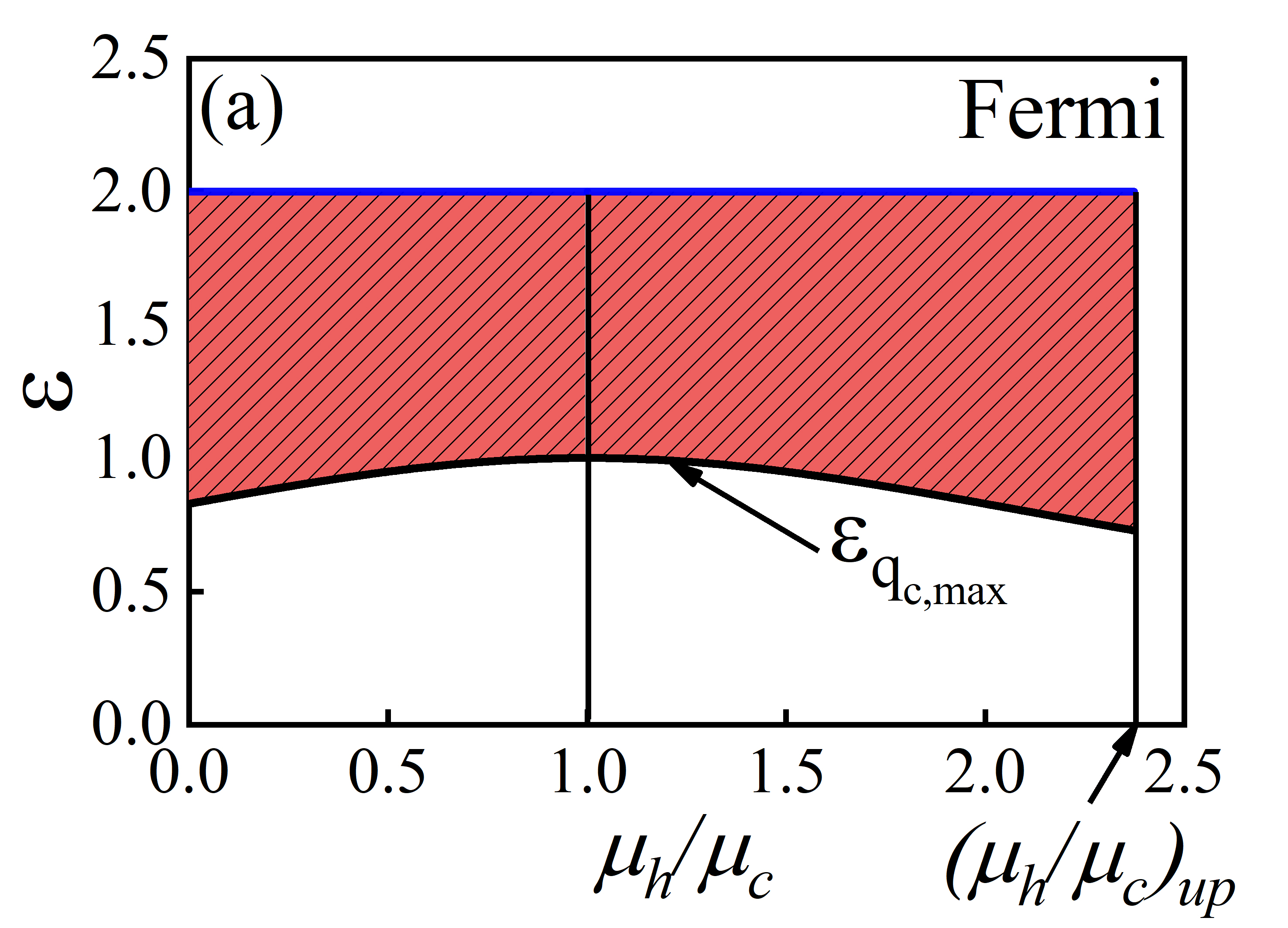}\includegraphics[scale=0.12]{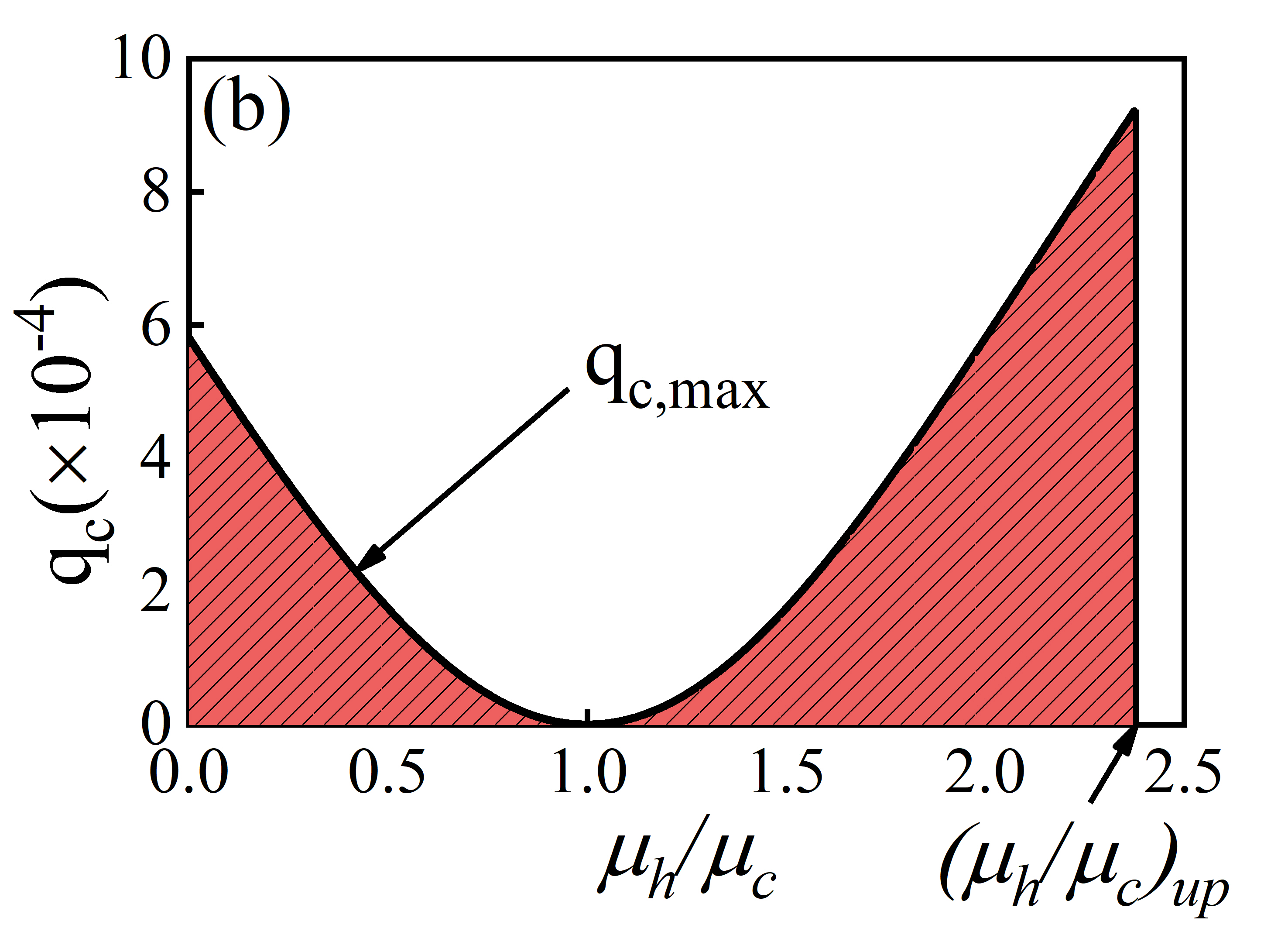}\includegraphics[scale=0.12]{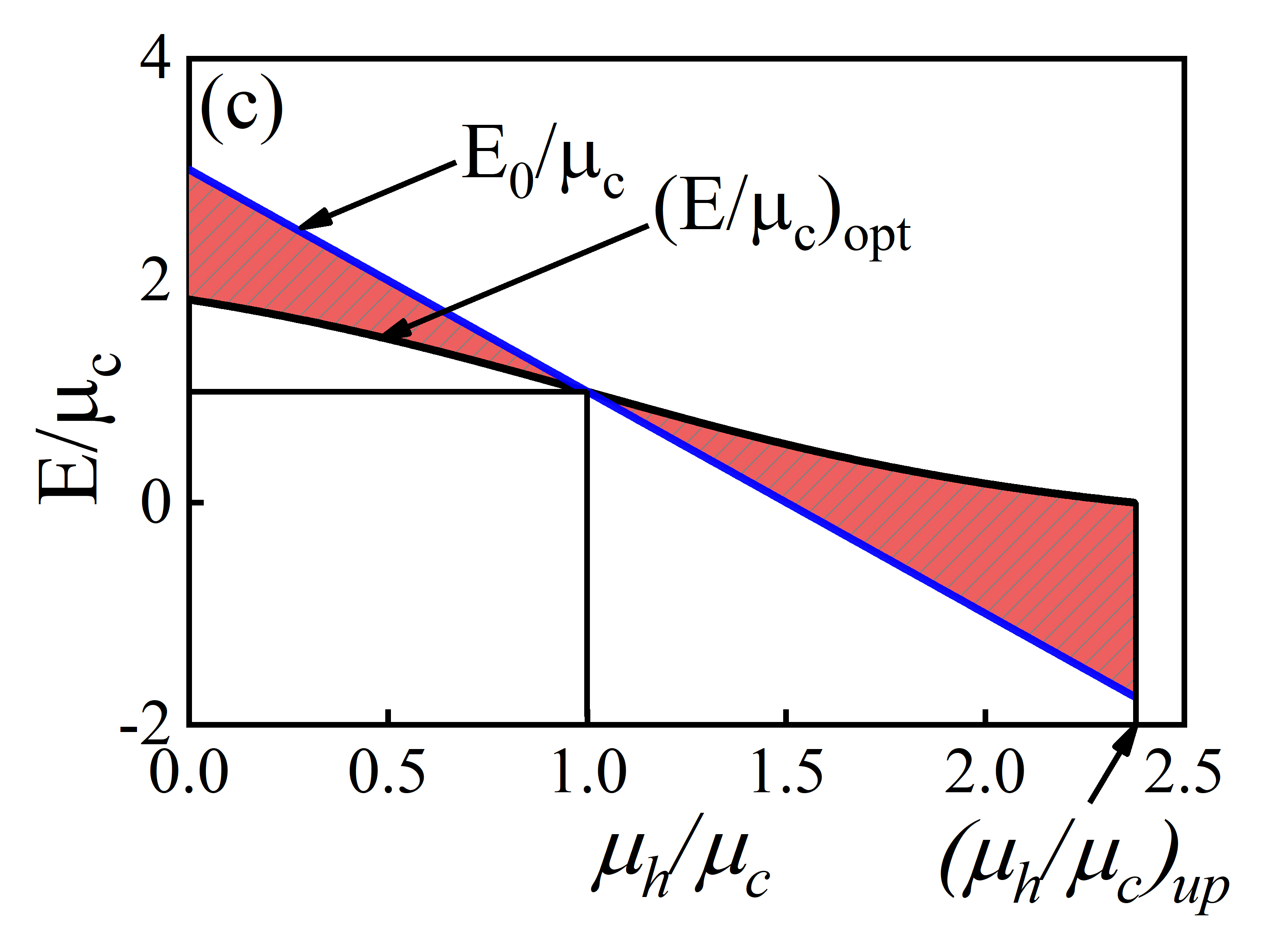}

\includegraphics[scale=0.12]{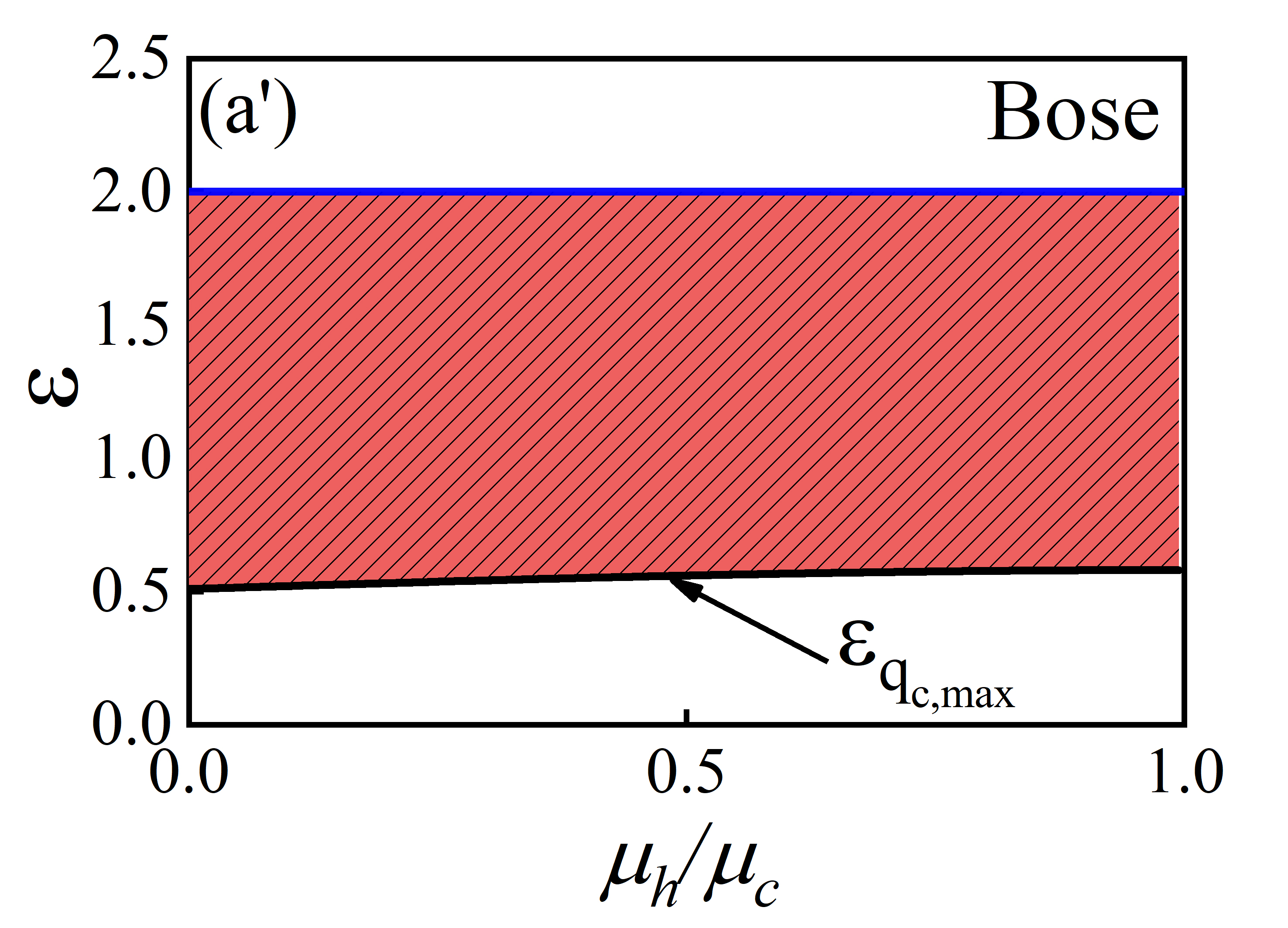}\includegraphics[scale=0.12]{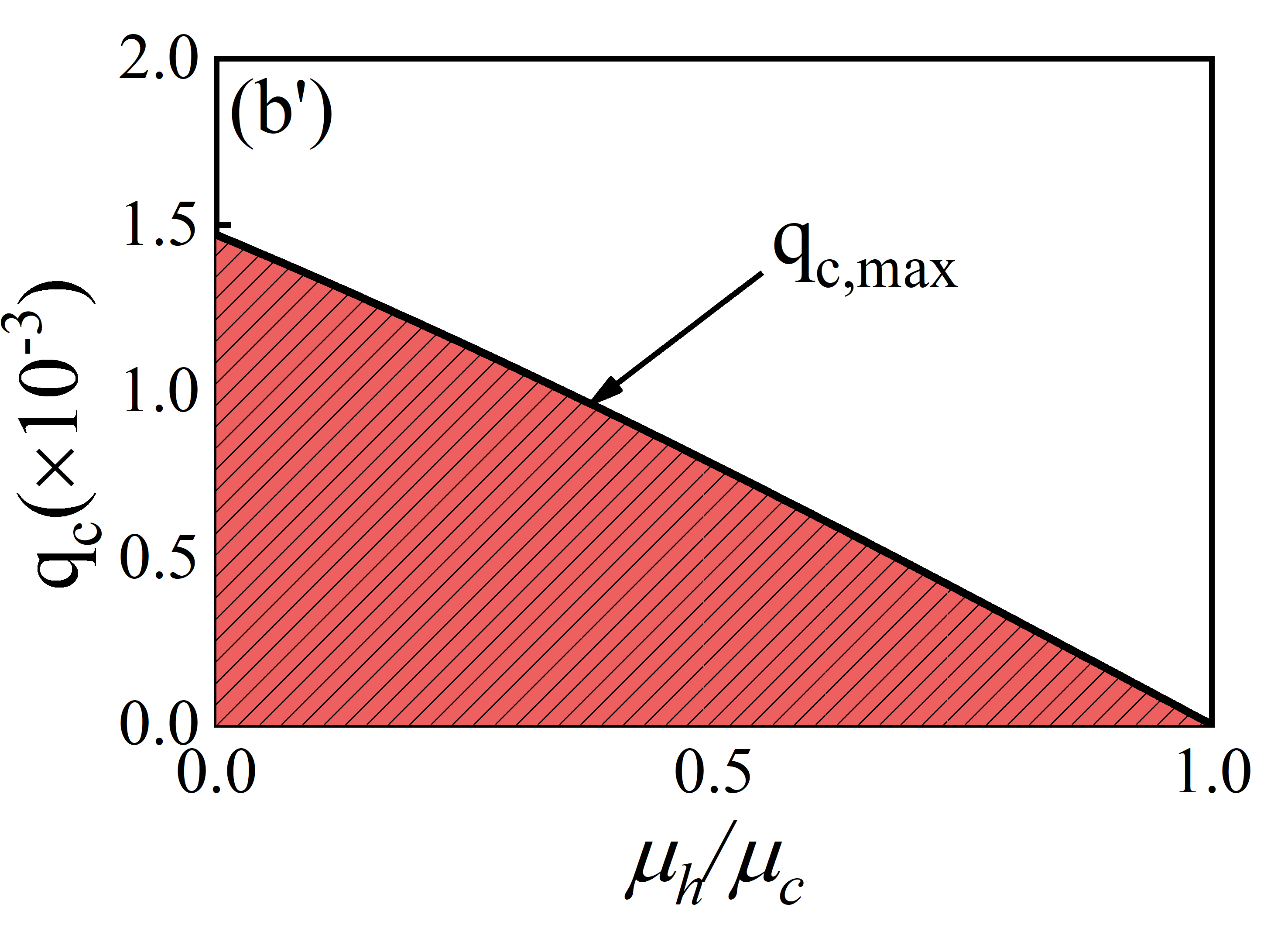}\includegraphics[scale=0.12]{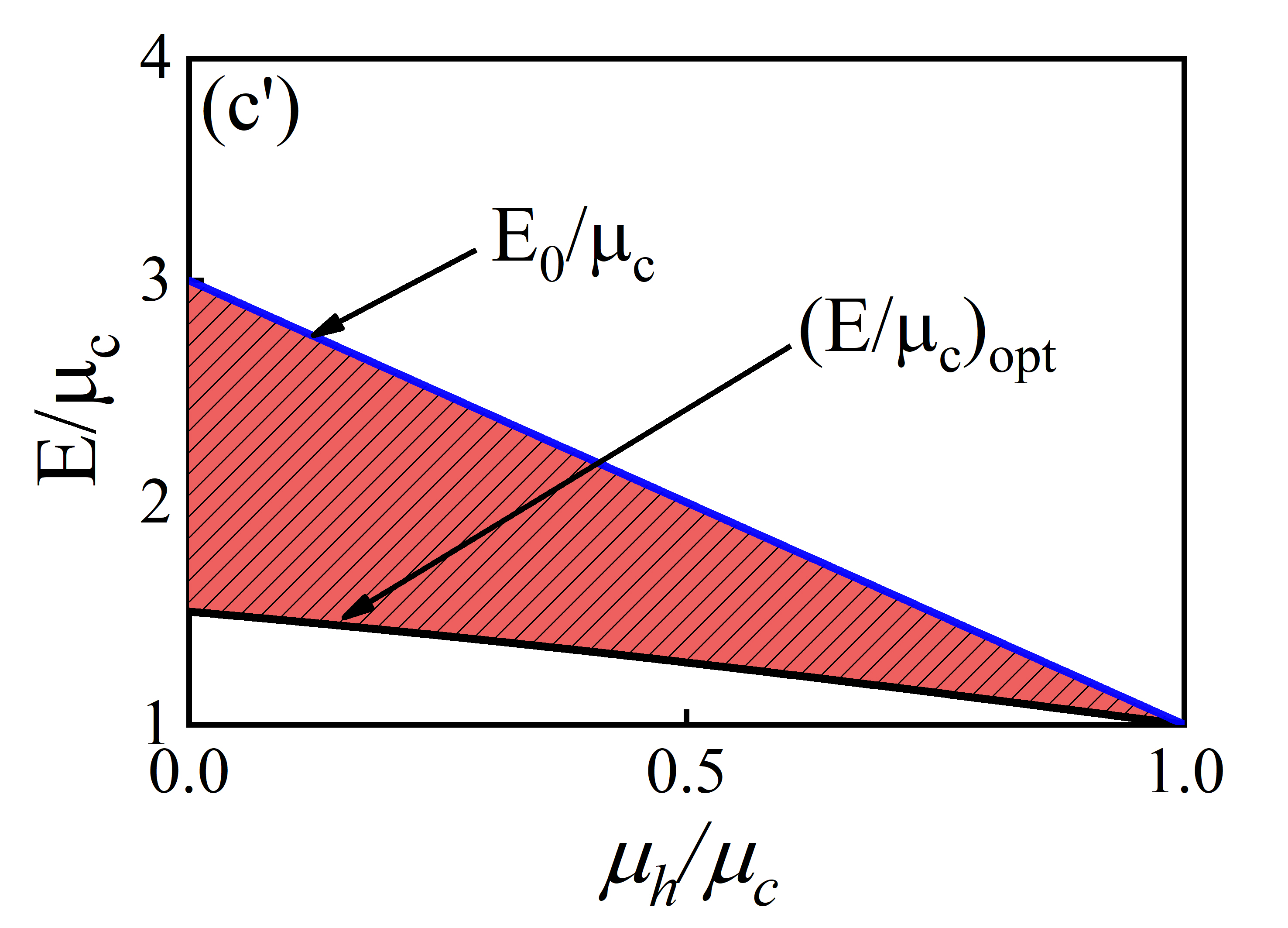}

\caption{The optimum regions of ($a$) {[}($a'$){]} the coefficient of performance
$\varepsilon$, ($b$) {[}($b'$){]} the cooling rate $q_{c}$, and
($c$) {[}($c'$){]} the energy level $E/\mu_{c}$ of the QD cooler
operating between two Fermi {[}Bose{]} reservoirs for differently
given values of $\mu_{h}/\mu_{c}$. The parameters $\mu_{c}=1.9$,
$T_{h}=3$, $T_{c}=2$, and $\Gamma_{h}=\Gamma_{c}=0.01$. \label{fig:8}}
\end{figure}

For a micro/nano cooler, one always wants to obtain a coefficient
of performance and a cooling rate as large as possible. According
to Figs. \ref{fig:4} and \ref{fig:7}, we can further determine the
optimum regions of $\varepsilon$, $q_{c}$ and $E/\mu_{c}$, as shown
in Fig. \ref{fig:8}. Form Fig. \ref{fig:8}, we can obtain the optimum
selection criteria of the key parameters of the QD cooler as
\begin{equation}
\begin{cases}
\begin{array}{c}
\varepsilon_{r}>\varepsilon\geq\varepsilon_{q_{c,max}}\\
0<q_{c}\geq q_{c,max}\\
E_{0}/\mu_{c}>E/\mu_{c}\geq(E/\mu_{c})_{opt}\\
E_{0}/\mu_{c}<E/\mu_{c}\leq(E/\mu_{c})_{opt}
\end{array} & \begin{array}{c}
\\
\\
(\mu_{h}/\mu_{c}<1),\\
(\mu_{h}/\mu_{c}>1)
\end{array}\end{cases}\tag{10}\label{eq:10}
\end{equation}
where $\varepsilon_{q_{c,max}}$ is the coefficient of performance
of the QD cooler at the maximum cooling rate $q_{c,max}$. Equation
(10) may provide some guidance for the experiment and development
of real micro/nano coolers.

\section{conclusions}

It has been proved that the single QD embedded between the two reservoirs
described by different statistical distribution functions can realize
the reverse flow and amplification of heat without externally driving
force and this seemingly paradoxical phenomenon does not violate the
laws of thermodynamics. It has been pointed out that the QD device
can work as a micro/nano cooler. The performance characteristics of
the cooler were revealed. The optimum operation regions of the cooler
were determined and the selection criteria of key parameters were
provided. The results obtained show that the proposed model has not
only guiding significance for the theoretical investigation of quantum
devices but also potential value for the practical application of
micro/nano devices.

\begin{acknowledgments}
This work is supported by the National Natural Science Foundation
(No. 12075197 and 11805159), People\textquoteright s Republic of China.
\end{acknowledgments}

\end{document}